\documentclass[superscriptaddress,prb,twocolumn]{revtex4}
\usepackage{graphicx}
\usepackage{amsmath,physics}
\usepackage{color}
\usepackage{ulem}
\usepackage{tikz}
\usepackage{pgfplots}
\pgfplotsset{
table/search path={csv_files/},
}
\usetikzlibrary{pgfplots.groupplots}
\usepackage{varwidth}
\usepackage[utf8]{inputenc}
\usepackage{hyperref}

\xdefinecolor{darkgreen}{RGB}{0,139,0}
\xdefinecolor{green}{RGB}{0,195,0}

\begin{document}
\title{Gate-free state preparation for fast variational quantum eigensolver simulations}
\author{Oinam Romesh Meitei}
\thanks{These two authors contributed equally}
\affiliation{Department of Chemistry, Virginia Tech, Blacksburg, VA 24061, USA}
\author{Bryan T. Gard}
\thanks{These two authors contributed equally}
\affiliation{Department of Physics, Virginia Tech, Blacksburg, VA 24061, USA}
\author{George S. Barron}
\affiliation{Department of Physics, Virginia Tech, Blacksburg, VA 24061, USA}
\author{David P. Pappas}
\affiliation{Physics Department, University of Colorado, Boulder, CO 80309}
\affiliation{National Institute of Standards and Technology, Boulder, CO 80305}
\author{Sophia E. Economou}
\affiliation{Department of Physics, Virginia Tech, Blacksburg, VA 24061, USA}
\author{Edwin Barnes}
\affiliation{Department of Physics, Virginia Tech, Blacksburg, VA 24061, USA}
\author{Nicholas J. Mayhall}
\email{nmayhall@vt.edu}
\affiliation{Department of Chemistry, Virginia Tech, Blacksburg, VA 24061, USA}
\begin{abstract}
The variational quantum eigensolver (VQE) is currently the flagship algorithm
for solving electronic structure problems on near-term quantum computers. This
hybrid quantum/classical algorithm involves implementing a sequence of
parameterized gates on quantum hardware to generate a target quantum state,
and then measuring the expectation value of the molecular Hamiltonian. Due to
finite coherence times and frequent gate errors, the number of gates that can
be implemented remains limited on current quantum
devices, preventing accurate applications to systems with significant
entanglement, such as strongly correlated molecules. In this work, we propose
an alternative algorithm (which we refer to as {\tt ctrl-VQE}) where the
quantum circuit used for state preparation is removed entirely and replaced by
a quantum control routine which variationally shapes a pulse to drive the
initial Hartree-Fock state to the full CI target state. As with VQE, the
objective function optimized is the expectation value of the qubit-mapped
molecular Hamiltonian. However, by removing the quantum circuit, the coherence
times required for state preparation can be drastically reduced by directly
optimizing the pulses. We demonstrate the potential of this method numerically
by directly optimizing pulse shapes which accurately model the dissociation
curves of the hydrogen molecule (covalent bond) and helium hydride ion (ionic
bond), and we compute the single point energy for LiH with four transmons.
\end{abstract}
\maketitle

\section{Introduction}
Molecular modeling stands in the juncture of key advances in many important
fields including and not limited to energy storage, novel material designs,
and drug discovery. For more than half a century, many seminal works have been
reported on the development of theories and methods to enable molecular
modeling with high accuracies. Approximate numerical methods which are built
on a single Slater determinant reference state, such as density functional
theory (DFT), perturbation theory, or coupled cluster, perform well when the
amount of electron correlation, ranges from minimal to moderate.\cite{}
However, for systems which are qualitatively governed by electron correlation
effects (strongly correlated systems), such approximate methods fail to be
sufficiently accurate. While alternative strategies exist, such as density
matrix renormalization group
(DMRG)\cite{whiteDensityMatrixFormulation1992,chanHighlyCorrelatedCalculations2002,schollwockDensitymatrixRenormalizationGroup2011}
or selected configurational interaction (SCI)
methods,\cite{Huron1973,Bender1969,Buenker1968,evangelistiConvergenceImprovedCIPSI1983,tubmanDeterministicAlternativeFull2016,schriberCommunicationAdaptiveConfiguration2016,holmesHeatBathConfigurationInteraction2016,levineCASSCFExtremelyLarge2020,abrahamSelectedConfigurationInteraction2020c,caffarelUsingCIPSINodes2016}
which can handle strong correlation, these approaches assume that the
correlations are either low-dimensional, or limited in scale. Currently, no
polynomially scaling classical algorithm exists which can solve for arbitrary
molecular ground states. 

High dimensionality of the wavefunction is ultimately the core reason for the
exponential cost of modeling electronic structure. Even before optimization,
simply storing the wavefunction on classical hardware quickly becomes a
bottleneck. This is because one typically represents the wavefunction in a
basis of ``classical'' states, or basis states which have a direct product
structure (Slater determinants, occupation number vectors, or even tensor
product states\cite{abrahamSelectedConfigurationInteraction2020c}). In this
classical basis, the exact (and generally entangled) state of the system is
represented as an exponentially\footnote{Or factorially, if a basis of Slater
  determinants is used to span a target projected spin subspace.} large vector 
of coefficients weighting the corresponding classical basis states. 

Quantum computing offers a radical departure from this computational strategy.
As quantum systems themselves, the state of a quantum processing unit (QPU) is
also a vector in a Hilbert space of identical dimension to the molecular
problem. This ability to perform a one-to-one mapping between vectors in the
Hilbert space containing the molecule's electronic wavefunction and those in
the state space accessible to a QPU means that with enough control over the
QPU, it should be possible to take the vector corresponding to the molecular
wavefunction and realize it on the QPU, avoiding altogether the requirement to
work with an exponentially large vector of coefficients. Once the QPU is
prepared into the state corresponding to the target molecule, any molecular
observable (energy, dipole moment, etc) can be obtained by measuring the
corresponding operator on the QPU. 

In order to turn this strategy into an algorithm, one needs a procedure for
realizing the target molecular wavefunction on the QPU. As the leading quantum
algorithm for molecular simulation on near-term devices, the Variational Quantum Eigensolver
(VQE)\cite{peruzzoVariationalEigenvalueSolver2014} provides an efficient
procedure for this purpose. In VQE, one defines a parameterized quantum
circuit comprised of tunable gates, and then optimizes these gates using the
variational principle, minimizing the energy of the molecular
Hamiltonian. This parameterized quantum circuit (referred to as an ansatz)
defines the variational flexibility (and thus the subspace reachable on the
QPU) of the algorithm. 

State-preparation circuits with more parameters generally have more
variational flexibility, but come with the cost of having deeper quantum
circuits and more difficult optimization. This cost can be significant. 
Current and near-term quantum computers are classified as noisy intermediate
scale quantum (NISQ) devices due to the presence of short coherence times,
system noise, and frequent gate
errors\cite{preskillQuantumComputingNISQ2018}. Because each gate has limited 
fidelity, the success probability for a sequence of gates decreases
exponentially with circuit depth. 
{ Even if gates could reach unit fidelity, 
the finite coherence times of a NISQ device still } limits the number of gates that one can
apply in a circuit which, in turn, limits the accuracy of the molecular VQE
simulation. Although VQE is relatively robust in the presence of noise and
errors in certain cases,\cite{Sharma_2020} the critical limitation preventing
larger scale experiments is the accurate implementation of deep circuits. The
goal of finding parameterized circuits which minimize the circuit depth and
maximize the accuracy has led to a number of approaches such as hardware
efficient ans\"atze,\cite{Kandala_2017} physically motivated fixed
ans\"atze,\cite{hugginsNonOrthogonalVariationalQuantum2019a,leeGeneralizedUnitaryCoupled2019,gardEfficientSymmetrypreservingState2020,Barron_2020,ryabinkinQubitCoupledclusterMethod2018}
and adaptive
ans\"atze.\cite{grimsleyAdaptiveVariationalAlgorithm2019,tangQubitADAPTVQEAdaptiveAlgorithm2020,ryabinkinIterativeQubitCoupled2020} 

{ Each of these mentioned approaches essentially aims to find the shortest quantum circuit for preparing a state to within an acceptable error. 
However, even if one succeeds at finding the ``best'' circuit, it is likely that its execution time will still exceed the coherence time of a NISQ device. 
In order to make meaningful progress toward larger problems, one can imagine three strategies:
(i) device improved error mitigation techniques, 
(ii) significantly increase the ratio of coherence to gate time, or 
(iii) simply abandon gates and directly optimize the electronic controls. 
Of these, the latter strategy appears to be the most feasible on near-term NISQ devices.
}


In this paper, we explore the possibility of performing gate-free VQE
simulations by replacing the parameterized quantum circuit with a direct
optimization of the laboratory-frame analogue control settings. In the
following sections, we argue that quantum control techniques are likely to be
better suited for fast VQE state preparation than the more conventional
circuit-based approaches on NISQ devices. We first provide a detailed overview
of circuit-based VQE, then introduce our proposed strategy, {\tt ctrl-VQE},
then discuss initial results {along with a strategy to avoid over-parameterization}, and finally compare to gate-based ans\"atze. 
Several technical aspects, numerical results, and additional discussion are
provided in the supplementary information.

\section{Method}

\subsection{Variational Quantum Eigensolver}\label{sec:vqe}
The VQE algorithm aims to leverage classical resources to reduce the circuit depth required for molecular simulation.
The algorithm finds optimal rotation angles for a parameterized quantum circuit of fixed depth by variationally minimizing the energy of a target molecule,
	which is obtained by repeated state preparation and measurement cycles.
In order to account for the distinguishability of qubits,
	we start by transforming the second quantized electronic Hamiltonian into an equivalent form involving non-local strings of Pauli spin operators, $\hat{o}_i$:
\begin{align}\label{eq:h_mol}
	\hat{H}^{\text{molecule}} &= \sum_{pq}h_{pq}\hat{p}^\dagger\hat{q} + \tfrac{1}{2}\sum_{pqrs}\ip{pq}{rs} \hat{p}^\dagger\hat{q}^\dagger\hat{s}\hat{r} \nonumber\\
	 &= \sum_i\hat{o}_i h_i,
\end{align}
	where $h_i$ is  a sum of  molecular one or two-electron integrals,
	and $\hat{p}$ operators are are fermionic annihilation operators. 
Several such transformations exist, such as the Jordan-Wigner,\cite{jordanwigner1928} Bravyi-Kitaev,\cite{bravyi2002} or parity transformation.\cite{parity}
The main steps in VQE are defined as follows:
\begin{enumerate}
	\item Precompute all $h_i$ values, and transform terms in the Hamiltonian operator into the desired qubit representation.
	\item Choose an  ansatz for the problem which defines the variables ($\vec{\theta}$) to optimize.
		Assuming one starts from the Hartree-Fock (HF) reference state, this involves predefining a parameterized circuit which
		provides enough variational flexibility to describe the molecule's electron correlation effects.
		Many ans\"atze have been proposed, several of which are variants of the original proposal using the Unitary Coupled-Cluster (UCC) ansatz.\cite{uccsd_bartlett,uccsd_vqe_alan}
	\item Choose an initial set of parameter values, $\vec{\theta} = \vec{\theta}_0$.
		These can be initialized to zero, chosen randomly,
		or if appropriate, chosen based on some classical precomputation such as using MP2 parameters to start a UCCSD optimization.
	\item Using current parameters, $\vec{\theta}$, repeatedly execute the circuit,
		each time performing an individual measurement of one of the operators, $\hat{o}_i$, in $\hat{H}$.\footnote{or a set of mutually commuting operators}
		After a sufficient number of circuit executions (shots), the
                averages of the resulting data converge to the expectation
                values of the operators 
		such that the average molecular energy can be obtained by
                multiplication with the one and two-electron integrals, 
		\begin{align}\label{eq:energy}
			E(\vec{\theta}) = \sum_i h_i \ev{\hat{o}_i}{\psi(\vec{\theta})}.
		\end{align}
	\item Check convergence. If the average energy has decreased by a small enough value determined to be converged, exit.
		If the energy is not yet converged, update $\vec{\theta}$, and return to step 4.
\end{enumerate}
Various approaches have been proposed, each differing in the details of
state-preparation,\cite{ostaszewskiQuantumCircuitStructure2019,tangQubitADAPTVQEAdaptiveAlgorithm2020,chivilikhinMoGVQEMultiobjectiveGenetic2020,jastrow_low_depth_Yuta,
  hugginsNonOrthogonalVariationalQuantum2019,leeGeneralizedUnitaryCoupled2018b}
	and ways to reduce the number of circuits required to compute expectation values \cite{babbush2018low, verteletskyi2019measurement, hugginsEfficientNoiseResilient2019, PhysRevA.101.062322}.

\subsection{Control Variational Quantum Eigensolver: {\tt ctrl-VQE}}\label{sec:C-VQE}
In this section, we present an alternative to the gate-based VQE algorithm,
replacing the parameterized state-preparation circuit with a parameterized
laboratory-frame pulse representation, which is optimized in an analogous
manner, but with the benefit of a much faster state preparation, opening up
the possibility of more accurate simulations on NISQ devices. 
All other aspects of VQE (i.e., measurement protocols) are
essentially the same. Using the molecular energy as the objective function to
minimize, the pulse parameters are optimized using the variational principle. 
This general strategy, which we refer to as control variational quantum
eigensolver ({\tt ctrl-VQE}), is outlined as follows: 

\begin{enumerate}
    \item As done in any regular VQE, compute the one- and two-electron integrals and transform
    the molecular Hamiltonian into a qubit representation, for example using Jordan-Wigner, Parity
    or Bravyi-Kitaev mappings.
		This defines the objective function to minimize, $~\langle \hat{H}^{\text{molecule}}\rangle$.
    \item  Define a fixed pulse representation (e.g. square pulses, sum of Gaussian pulses, etc.).
	    Parametrize the chosen pulse representation, and initialize parameters.
    \item Choose an initial state for the qubit(s) system. HF is a good choice
    for the molecular problems studied here. Controls are assumed to be in the form of direct drives on each qubit.
    \item Measure the expectation value of the objective function $ \langle \hat{H}^{\text{molecule}} \rangle$ on the
    quantum device.
    \item Using a classical optimization routine, determine the pulse parameters for the
    next measurement on the quantum device.
    \item Repeat until a target convergence threshold is met.
    If the chosen parameterized pulse can fully specify the target
    Hilbert-space, then the optimal pulse finds the minimum energy state.  
\end{enumerate}

We note here that the optimization used in this work excludes the total
pulse duration. This is fixed throughout the optimization routine, and only the pulse parameters
such as the amplitudes or the frequencies are directly optimized.
As such, the total pulse time enters the algorithm as a ``hyper-parameter'',
which can be optimized in an outerloop if desired.
In the square pulses considered in  this work, the time segments are also
optimized. Unless stated otherwise, optimal pulses 
correspond to pulse shapes that are optimal with a given fixed total pulse duration.

Unlike universal quantum computing algorithms, {\tt ctrl-VQE} occurs at the
hardware-level, and any simulation must refer to a specific physical platform.
For this work, we choose a well-established transmon platform with the
following 1D device Hamiltonian\cite{kochChargeinsensitiveQubitDesign2007}:
\begin{align}\label{eq:transmons}
	\hat{H}_D =&\sum_{k=1}^N \left(\omega_k \hat{a}_k^\dagger\hat{a}_k-\frac{\delta_k}{2}\hat{a}_k^\dagger\hat{a}_k^\dagger\hat{a}_k\hat{a}_k \right)\\
				&+ \sum_{\ev{kl}}g(\hat{a}_k^\dagger\hat{a}_l+\hat{a}_l^\dagger\hat{a}_k). \nonumber
\end{align}
where $\ev{kl}$ indicates that the summation is over pairs of coupled
qubits. Here, we assume that the qubits are arranged in a closed linear chain and
have always-on direct couplings,
where $\hat{a}_k$ is the bosonic annihilation operator for the $k^{\textrm{th}}$ transmon,
and $\omega_k$, $\delta_k$, and $g$ are the resonant frequency, anharmonicity,
and constant coupling rate, respectively. 
Furthermore, each transmon in Eq.~(\ref{eq:transmons}) formally supports an infinite number of states.
However, in our simulations we necessarily approximate this system by a finite
number of levels (three unless otherwise stated) per transmon. 
We tested the accuracy of this by adding more levels and found that the
results did not change significantly. 
The parameters used in this work are chosen to be  typical of values found in current superconducting transmons, 
and are provided in Table \ref{table:system_parameters}.
We find that our results below do not qualitatively depend on the frequency difference between the qubits,
	and in the SI we provide a comparison of this current device to one with a larger detuning between the transmons.
In order to drive the device, we apply a time-dependent field, with separate controls on each qubit
such that within the rotating wave approximation, the control Hamiltonian is expressed as:
\begin{align}
	\hat{H}_C =&\sum_{k=1}^N \Omega_k(t)(e^{i\nu_k t} \hat{a}_k+e^{-i\nu_k t}\hat{a}_k^\dagger).
\end{align}
where $\Omega_k(t)$ is the real-valued, time-dependent amplitude of the drive, and $\nu_k$ is the frequency of the field.
The system therefore evolves under the total Hamiltonian:
\begin{align}
	\hat{H} = \hat{H}_D + \hat{H}_C(t, \Omega_n(t), \nu_n).
\end{align}

By moving to the interacting frame, the final ansatz has the following form:
\begin{align}\label{eq:ansatz}
	\ket{\psi^{\text{trial}}(\Omega_n(t), \nu_n)} =& \mathcal{T}e^{-i\int_0^Tdt \hat{\tilde{H}}_C(t, \Omega_n(t), \nu_n)}\ket{\psi_0}.
\end{align}
where $\ket{\psi_0}$ is the VQE reference state (e.g., the HF state),
$\{\Omega_n(t),\nu_n\}$ are the variational parameters for the ansatz,
$T$ is the total pulse time, and $\mathcal{T}$ is the time-ordering operator.
Note that although the control Hamiltonian above only has single-qubit terms,
the device itself has inter-qubit couplings (Eq. \ref{eq:transmons} with strength $g$),
which create an entangling control Hamiltonian in the interacting frame:
\begin{align}
	\hat{\tilde{H}}(t)_C = e^{i\hat{H}_Dt}\hat{H}_C(t)e^{-i\hat{H}_Dt}.
\end{align}
As such, the coupling strength $g$ is ultimately responsible for describing electron correlation in
the target molecule.

Using this ansatz in Eq. \ref{eq:ansatz},
the energy to be minimized in the {\tt ctrl-VQE} objective function is,
\begin{align}
	E(\Omega_n(t), \nu_n) = \bra{\psi^{\text{trial'}}}\hat{H}^{\text{molecule}}\ket{\psi^{\text{trial'}}}.
\end{align}
where $\ket{\psi^{\text{trial'}}}$ is just the $\ket{\psi^{\text{trial}}}$ state above, projected onto the 
computational basis and normalized.
Note that an unnormalized state can also be used; this yields similar results, as shown in the SI.

The ansatz above is completely (and solely) determined by the device and controls,
granting enormous flexibility to the ansatz.
In fact, any digital quantum circuit ansatz can be compiled into the form above.
As such, the ansatz in Eq. \ref{eq:ansatz} does not intrinsically possess any
limitation on its potential accuracy 
	beyond the fundamental limitations imposed by quantum speed limits.\cite{QSL_2017}
However, this additional flexibility can make optimization more difficult.
\begin{figure} 
  \centering
  \includegraphics[width=0.5\linewidth]{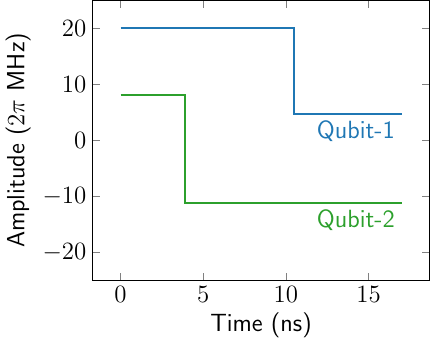}
	\caption{Illustration of a simple square pulse with two time segments.    }
   \label{fig:pulse_illus}
\end{figure}
In this work, we have chosen to impose simple constraints on the form of
$\Omega_n(t)$ {to minimize the number of optimization parameters and simplify
experimental implementation.} 
We have considered two examples: i) piecewise-constant pulses, and ii) sum of Gaussian pulses.
Because these two examples yield similar results, we present only the square
pulse data in the main text, 
and provide the Gaussian pulse data in the SI.
{An example of a pair of two-time-segment square pulses for a two
  transmon system is shown in Figure \ref{fig:pulse_illus}.
  With a single time segment square pulse, the amplitude
  is taken as constant throughout the pulse duration.
  For $n$ time segments, the
parameterized square pulse is given by,
\begin{equation}
  \Omega_k(t)=
    \begin{cases}
      c_1 & 0\leq t < t_1\\
      c_2 & t_1\leq t < t_2\\
      \vdots\\
      c_n & t_{n-1}\leq t < T \quad,
    \end{cases}
\end{equation}
where $c_i$ are amplitudes of the pulse, $t_i$ are the switching times,
and $T$ is the total pulse duration.
The amplitudes are constrained to $\pm$20 MHz unless otherwise mentioned; this is a typical amplitude for RF control signals in superconducting qubit
systems \cite{hyperbolic_secant_pulse_pappas}}
Each transmon drive term also has a frequency modulation of the form
$
\mbox{exp}\left(i~\nu_k t\right),
$
with a driving frequency $\nu_k$ constrained to $(\omega_k-2\pi \text{ GHz})\leq \nu_k
\leq(\omega_k+2\pi \text{ GHz})$, where $\omega_k,\nu_k$ are in units of
$2\pi~\mbox{GHz}$. Therefore, the pulse parameters that will be optimized
include $c_j$, $t_j$ and $\nu_k$. With $N$ transmons and $n$ 
square pulses on each transmon, we then have $2Nn$ parameters to
optimize. The pulse parameter optimizations are performed using
l-BFGS-b.

We note that a very recent preprint uses similar quantum control considerations to improve VQE.\cite{choquetteQuantumoptimalcontrolinspiredAnsatzVariational2020}
However, in that work, these considerations are used only to define an improved gate-based ansatz,
unlike the direct variational pulse shaping described in this work.
Our work is also somewhat related to the paper from Chamon and coworkers (Ref. \onlinecite{yangOptimizingVariationalQuantum2017}) 
	in which they find that the optimal pulse for preparing the ground state of a Hamiltonian is of a bang-bang form.
Although both {\tt ctrl-VQE} and the work of Ref. \onlinecite{yangOptimizingVariationalQuantum2017} both combine variational quantum algorithms with quantum control,
	our effort aims to find ground states of non-diagonal Hamiltonians (such as the molecular Hamiltonian in Eq. \ref{eq:h_mol}),
	preventing the efficient implementation of the time-evolution operator needed for realizing the bang-bang protocol. 
Because our control Hamiltonian is not determined by our problem Hamiltonian (Eq. \ref{eq:h_mol}),
	we should not expect such bang-bang protocols to be optimal.

\section{Computational Details}\label{sec:comp_detail}
{The numerical results presented in this work were generated with
  a locally
developed program: {\tt CtrlQ}\footnote{The source code for the program is available on
  Github, codebase: \url{https://github.com/oimeitei/ctrlq} under a Apache
  License 2.0.}.
Functionalities from Qiskit\cite{Qiskit} and Qutip\cite{qutip} were used to
check and compare the results presented in this work.}
Molecular integrals were generated using PySCF\cite{pyscf}, and an STO-3G basis set was used throughout.
Below, we consider simulations that involve either two-transmon or four-transmon devices depending on the molecule.
The parameters $\omega$, $\delta$ and $g$
used in the simulations
are explicitly given in Table
\ref{table:system_parameters}. 
To demonstrate qualitative insensitivity to the device parameters, 
we provide some results using a device with a larger detuning in the SI.

{
Although the simulations presented in the following sections use highly restricted forms of control pulses, 
our in-house code ({\tt CtrlQ}) 
supports various forms of pulses.  
In fact, we have implemented efficient analytic gradients of the molecular
energy with respect to pulse amplitudes (see SI),
making it possible to optimize pulses with arbitrary shapes.
As mentioned in the previous section, 
here we focus primarily on simple pulse waveforms with few parameters to simplify experimental implementations.}

\begin{table}[] 
    \centering
    \begin{tabular}{l@{\hspace{6mm}}c@{\hspace{4mm}}c@{\hspace{4mm}}c@{\hspace{4mm}}c}
      \hline\hline
      & \multicolumn{4}{c}{Transmon}\\
      \hline
      &1&2&3&4\\
      \hline      
      $\omega$ &4.8080& 4.8333 & 4.9400 & 4.7960\\
      $\delta$ &0.3102 & 0.2916 & 0.3302 & 0.2616\\  
      \hline
      \noalign{\vskip1mm}
      &1 $\leftrightarrow{}$2&2 $\leftrightarrow{}$3&3 $\leftrightarrow{}$4&4
      $\leftrightarrow{}$1\\
      \hline
      $g$ & 0.01831 & 0.02131 & 0.01931 & 0.02031 \\
      \hline\hline
    \end{tabular}
    \caption{Device parameters appearing in Eq. \ref{eq:transmons} for a
      four-transmon system. For the simulations of H$_2$ and HeH$^+$, which
      only involve two transmons, device parameters corresponding to
      transmons 1 and 2 were used. All units are in $2\pi~\mbox{GHz}$.}
    \label{table:system_parameters} 
\end{table}


\section{Results and Discussion}\label{sec:results}
In the following subsections,
	we explore the ability of {\tt ctrl-VQE} to reproduce Full Configuration Interaction (FCI) (exact diagonalization)
	bond dissociation energy curves for two small example diatomic molecules
	with complementary electron correlation profiles.
We then analyze the dependence of the pulse time on the molecule's electron correlation,
	followed by a comparison to a representative gate-based circuit.
After reporting the new results,
	we discuss comparisons with related quantum algorithms.

\begin{figure} 
  \centering
  \includegraphics[width=0.85\linewidth]{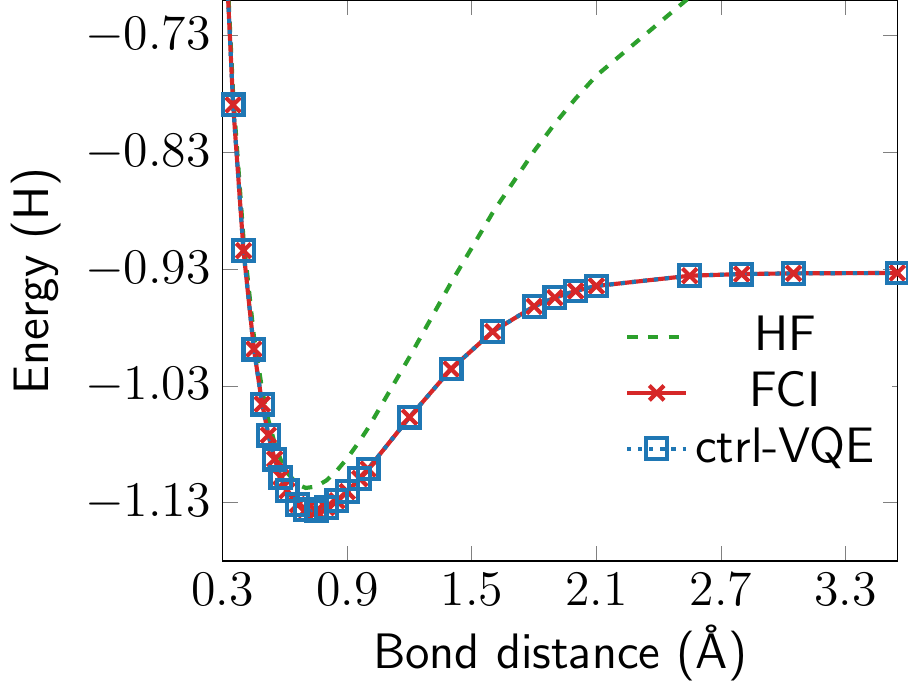}
	\caption{Bond dissociation curve of H$_2$ molecule. {\tt ctrl-VQE}
          energies are
    computed using square pulses with two time segments.}
  \label{fig:H2_sq_sd_tb}
\end{figure}

\begin{figure}
  \centering
  \includegraphics[width=0.85\linewidth]{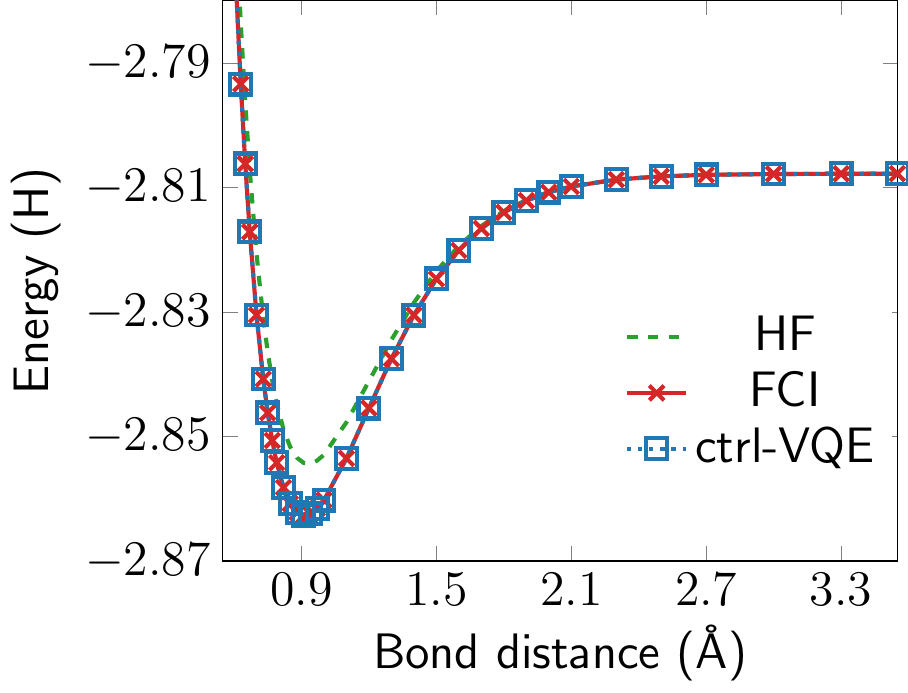}
	\caption{Bond dissociation curve of HeH$^+$ molecule. {\tt ctrl-VQE}
          energies are
        computed using square pulses with two time segments.}
      \label{fig:HeH_sq_sd_tb}
\end{figure}

\subsection{Dissociation of diatomic molecules}
Here, we demonstrate the performance of our approach by computing the ground state molecular
electronic energy along the bond dissociation of  the H$_2$ molecule and the HeH$^+$ molecular ion.
Although small, these molecules
have been used as benchmarks for quantum algorithms in recent years.\cite{wangQuantumSimulationHelium2015,Shen2017,Kandala_2017,Colless2018,mccaskey2019quantum,gardEfficientSymmetrypreservingState2020,Barron_2020,ishing_kais,trapedIon_Roos,mcqc_takui,Ritter_2019,symm_verf_vqe,compchem_qc_lianos}
These two molecular examples are chosen because they exhibit different
behaviors during bond dissocation in terms of their electronic structure.
As two-orbital, two-electron problems, these would naturally be modeled on
four qubits following the commonly used Jordan-Wigner transformation.\cite{jordanwigner1928}
However, to make the pulse simulations more computationally efficient, we have used the parity mapping instead
in which two qubits are diagonal and can be removed from consideration.
Details are given in Ref. \cite{Bravyi_2017}, and we use the implementation for mapping and pruning in Qiskit software.\cite{Qiskit}

As a prototypical example of a homolytic dissociation,
\begin{align}\label{eq:H2rxn}
	\text{H}_2 & \rightarrow\text{H}^\bullet+\text{H}^\bullet\\
	\text{Dynamic Correlation } &\rightarrow 	\text{Static Correlation }\nonumber
\end{align}
as H$_2$ dissociates, the ground state moves from being dominated by a single Slater determinant to
increasingly multiconfigurational, due to the shrinking HOMO-LUMO gap.
As a result, the accuracy of a mean-field treatment (e.g., HF) diminishes as the bond length is stretched.

In stark contrast, HeH$^+$ becomes easier to model as the bond distance increases.
The reason is that, being the strongest acid and, interestingly, the first molecule formed in the universe,\cite{gustenAstrophysicalDetectionHelium2019}
dissociation is a heterolytic deprotonation,
\begin{align}\label{eq:HeHrxn}
	\text{HeH}^+ & \rightarrow\text{He}+\text{H}^+\\
	\text{Dynamic Correlation } &\rightarrow 	\text{No Correlation*}\nonumber
\end{align}
such that both the reactants and the products are both closed shell and well represented by a single Slater determinant.
If fact, in a minimal basis set (i.e., the STO-3G basis set used in this paper),
	the products have exactly zero electron correlation energy
	because the H$^+$ ion has no electrons, while the He atom has no empty virtual orbitals into which electrons can be promoted.
As a result, HF becomes exact at dissociation.
Of course, in larger basis sets, the He atom would have some dynamic correlation, hence the * in Eq. \ref{eq:HeHrxn}.

{
The dissociation curves of the H$_2$ and HeH$^+$ molecular systems produced with
{\tt ctrl-VQE} using the simple square pulses are presented in Figs.
\ref{fig:H2_sq_sd_tb} and \ref{fig:HeH_sq_sd_tb} respectively.
Consistent with the physical descriptions above,
the HF state moves quickly away from the FCI ground state with bond dissociation
for H$_2$, while the opposite is true for HeH$^+$, where the HF state gradually converges to
the exact FCI ground state with increasing bond distance.
{\tt ctrl-VQE} reproduces the FCI bond dissociation curve of H$_2$ and HeH$^+$ with
high accuracies.}
The maximum difference from the FCI energy along the dissociation curve is $.03$ mHa
for both  H$_2$ and HeH$^+$,
and the average error is $.002$ mH in both the cases.
More detailed information including the pulse parameters, characteristics, and
molecular energies along the dissociation curves are provided in the SI.
The resulting states have a good overlap with the exact FCI ground states
(99\% in all cases).
For H$_2$, it is possible for the
overlap to deviate at long bond distances due to degenerate singlet and triplet states.
As the bond distance increases, the singlet ground state of H$_2$ becomes degenerate with
the lowest triplet state,
making it possible to generate a superposition of a singlet and triplet.
It is possible to converge to the exact ground singlet state by supplementing
the objective function with a penalty term proportional to the total
spin operator (see SI for details).

\begin{figure} 
  \centering
  \includegraphics[width=0.75\linewidth]{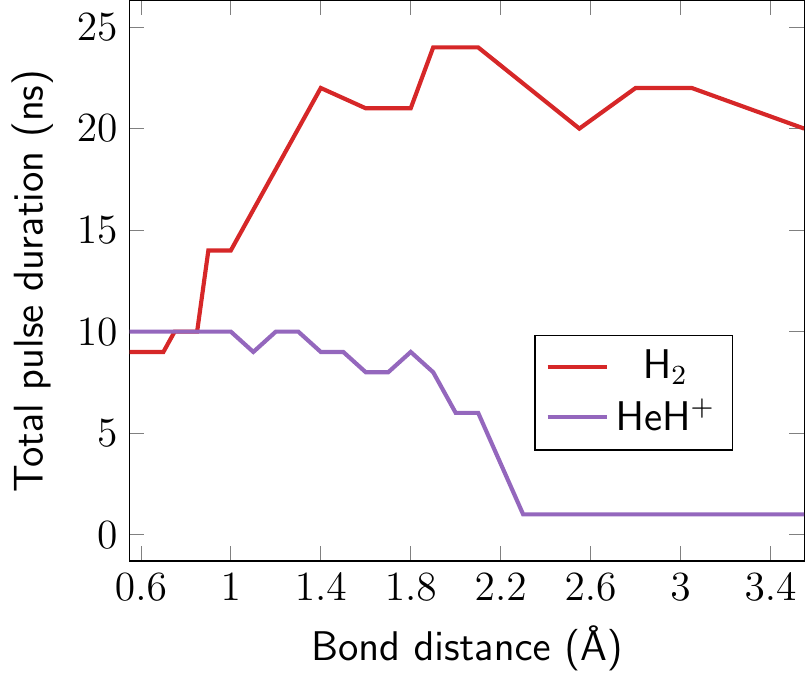}
  \caption{Total pulse duration as a function of  bond
    distance of H$_2$ and HeH$^+$ molecular systems. Square pulses
	were used to generate the {\tt ctrl-VQE} molecular energies in Figs.
    \ref{fig:H2_sq_sd_tb} and \ref{fig:HeH_sq_sd_tb}. The shortest pulse
    duration at each geometry is plotted here.
        }\label{fig:bond_duration}
\end{figure}

\subsection{Effect of electron correlation on pulse duration}
In any VQE, one is typically interested in finding a useful balance between accuracy (needing deep circuits)
and noise (needing shallow circuits).
As such, molecular simulations of strongly correlated molecules are intrinsically more difficult
	as deeper circuits are required, which increases problems from noise and gate errors.
An analogous balance is targeted in {\tt ctrl-VQE},
	where one would hope to obtain sufficiently accurate results with as short of a pulse time as possible.
Because entanglement cannot be created instantly,\cite{QSL_2017}
	we expect molecules with strong correlation to require longer pulses than simpler molecules.

In order to examine this relationship, in Fig. \ref{fig:bond_duration} we plot
the duration of the shortest pulses we were able to obtain
at each molecular geometry, with H$_2$ (HeH$^+$) shown in red (purple).
Referring back to Eqs. \ref{eq:H2rxn} and \ref{eq:HeHrxn},
	one would expect that as the bond distance is increased,
	H$_2$, needing more entanglement, would in turn require increasing pulse durations,
	whereas pulses for HeH$^+$ would get increasingly fast as the bond is stretched.
This trend is indeed observed.

The total pulse duration significantly decreases for the dissociation of
the HeH$^+$ molecular ion, whereas it significantly increases for the H$_2$
molecule. Note that the initial state and the final target state for
HeH$^+$ become degenerate with increasing bond distance (above 2.0 \AA).
Thus, the Hartree Fock state ($\ket{01}$) is a good approximation to the exact
FCI ground state, and only a slight modification to the initial state is
required to well approximate the ground state. With {\tt ctrl-VQE}, the total pulse
duration at the far bond distances of HeH$^+$ are only 1.0 ns. This directly
reflects the efficiency of the method presented in this work. In a gate-based
compilation method, generally one would still construct an ansatz comprised of
costly two-qubit gates, even though the target state is very nearly 
approximated by the initial state.

The total pulse duration at the dissociation limit for H$_2$ is significantly
longer than near the equilibrium bond distance. Although the HF molecular
energy increases monotonically beyond the equilibrium point,
the same is not observed for the pulse duration. This is suggestive of the
pulse durations reflecting the different dynamics along the bond dissociation
of the two molecular systems (see SI for a more detailed analysis).

\begin{figure}
  \centering
  \includegraphics[width=0.7\linewidth]{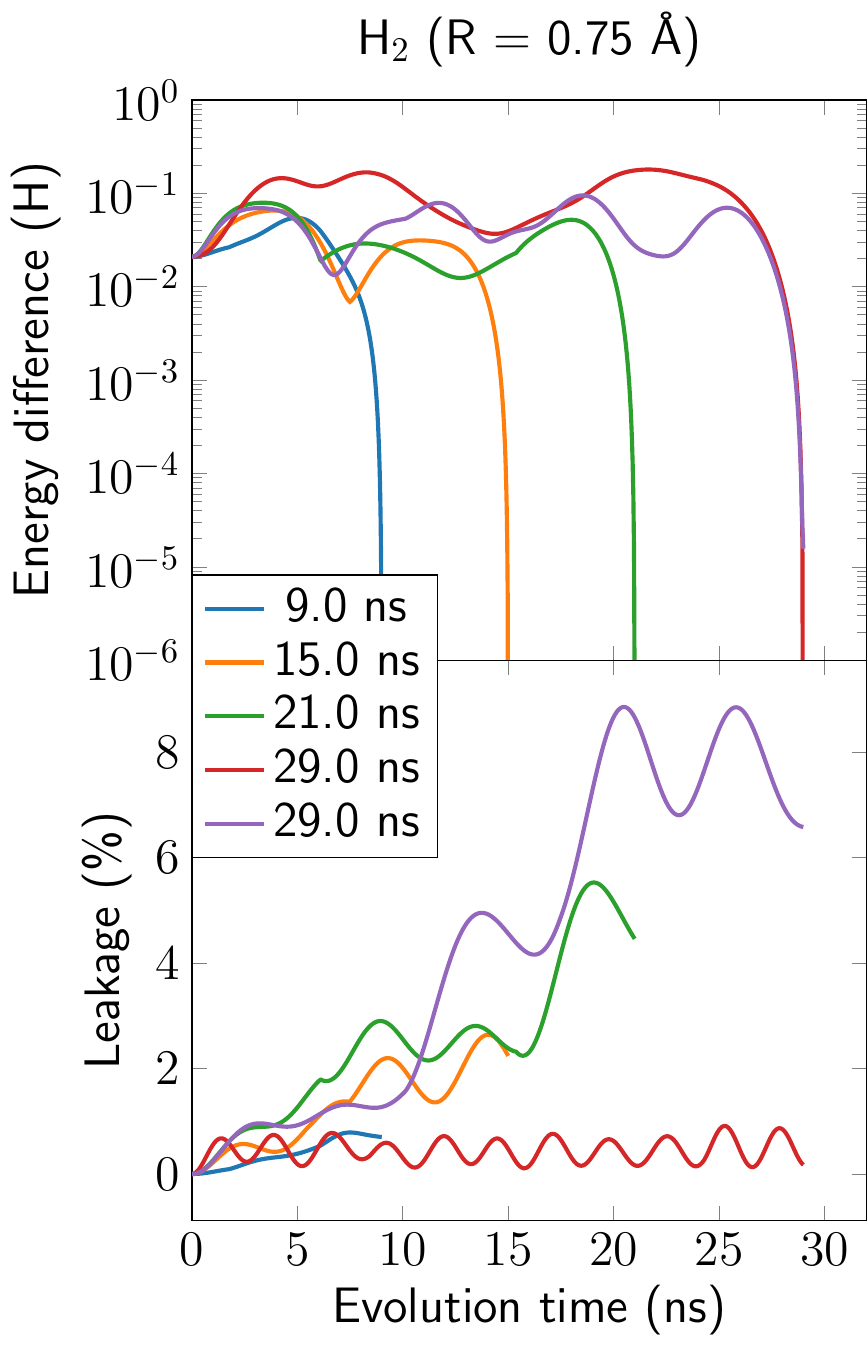}
  \caption{
    Energy difference between ctrl-VQE and FCI (top) and leakage (bottom) along the
    time evolution steps with optimal pulses of different durations for H$_2$
    with a bond distance of 0.75 \AA.
	The red and purple lines with $T=29$ ns are distinct solutions to the same optimization. 
    Optimized pulse parameters are provided in the SI.}\label{fig:error_leakage_time_h2}
\end{figure}

\begin{figure}
  \centering
  \includegraphics[width=0.7\linewidth]{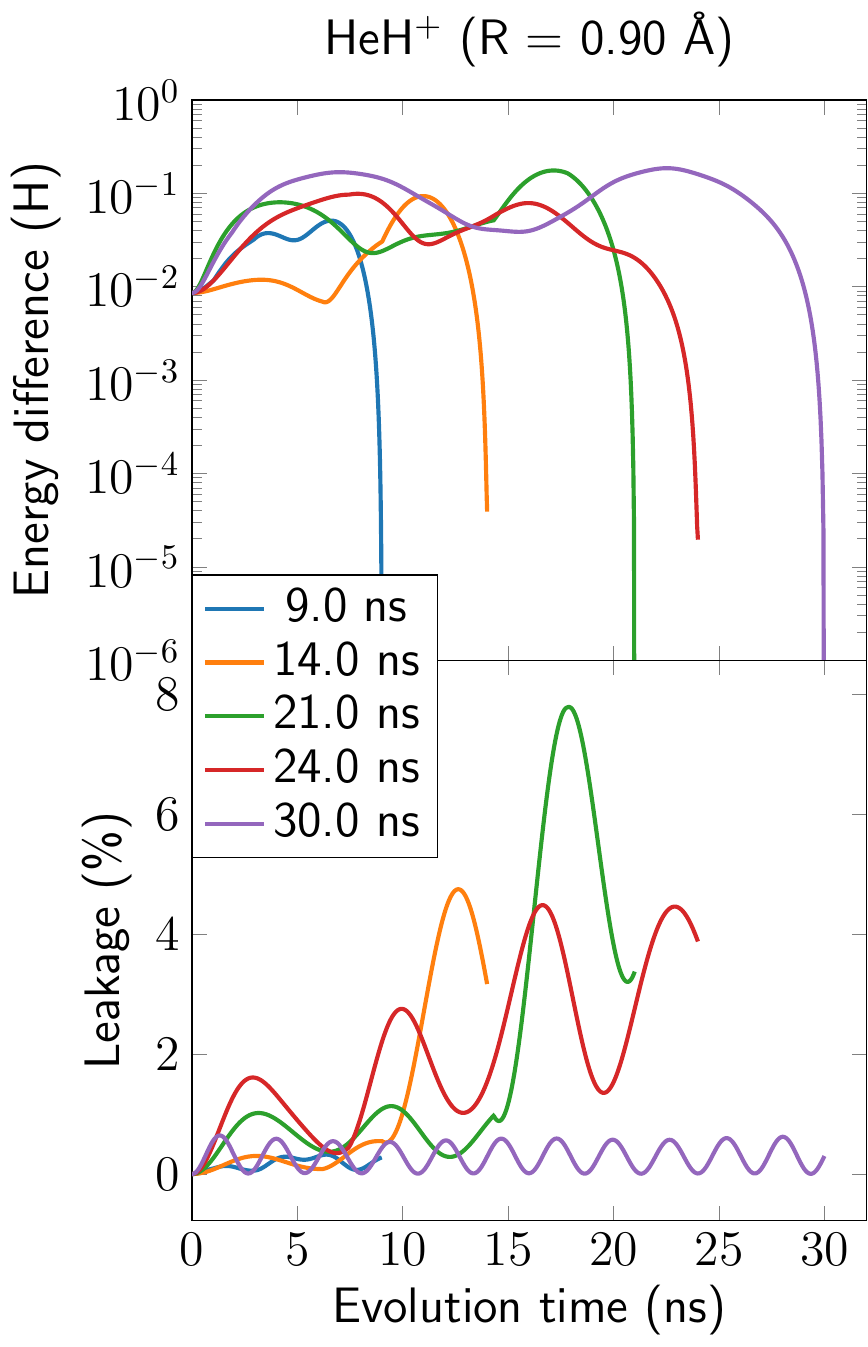}
  \caption{
    Energy difference between ctrl-VQE and FCI (top) and leakage (bottom) along the
    time evolution steps with optimal pulses of different durations for HeH$^+$
    with a bond distance of 0.90 \AA.
    Optimized pulse parameters are provided in the SI.}\label{fig:error_leakage_time_heh}
\end{figure}

\subsection{Leakage to higher energy states}
In Figs. \ref{fig:error_leakage_time_h2} and \ref{fig:error_leakage_time_heh}, 
we show the energy and leakage as a function of evolution time
for various choices of the hyper-parameter, $T$, (the total evolution time).
In each of the two figures, the top panel displays the behaviour of the
molecular energy as the state evolves from the reference Hartree-Fock state to the converged trial state,
and the bottom panel displays the total population outside of the $\ket{0}$ and $\ket{1}$ computational states for each qubit (leakage). 

We see that for each value of $T$, the molecular energy of both H$_2$ and HeH$^+$ never decreases toward the exact energy monotonically,
but rather increases initially, and then either decreases or oscillates before rapidly converging to the exact energy.
While each of the optimized pulses do indeed generate suitable transmon
dynamics which accurately produce the molecular ground state at the specified time $T$,
the different pulses are not each equally favorable. 
From the bottom panels of Figs. \ref{fig:error_leakage_time_h2} and \ref{fig:error_leakage_time_heh}, we see that some pulses create much more leakage than other pulses. 
High levels of leakage will likely require a larger number of shots (pulse executions) to get precise determinations of the expectation values,
since a higher portion of the shots end up in excited states which are either discarded, or collected for normalizing the results. 
However, with leakage of only 10\%, we anticipate only needing a commensurate increase in the shot count to compensate for post-selection,
and so we do not expect this to be a fundamental limitation of the approach. 
Alternatively, one can use an unnormalized energy in the objective function. 
This approach naturally constrains the solutions to simultaneously minimize leakage, 
	as any leakage necessarily penalizes the associated pulse.
Results using an unnormalized energy in the objective function are comparable to those shown here, 
	and are given in the SI.

\begin{figure}
    \centering
  \includegraphics[width=0.75\linewidth]{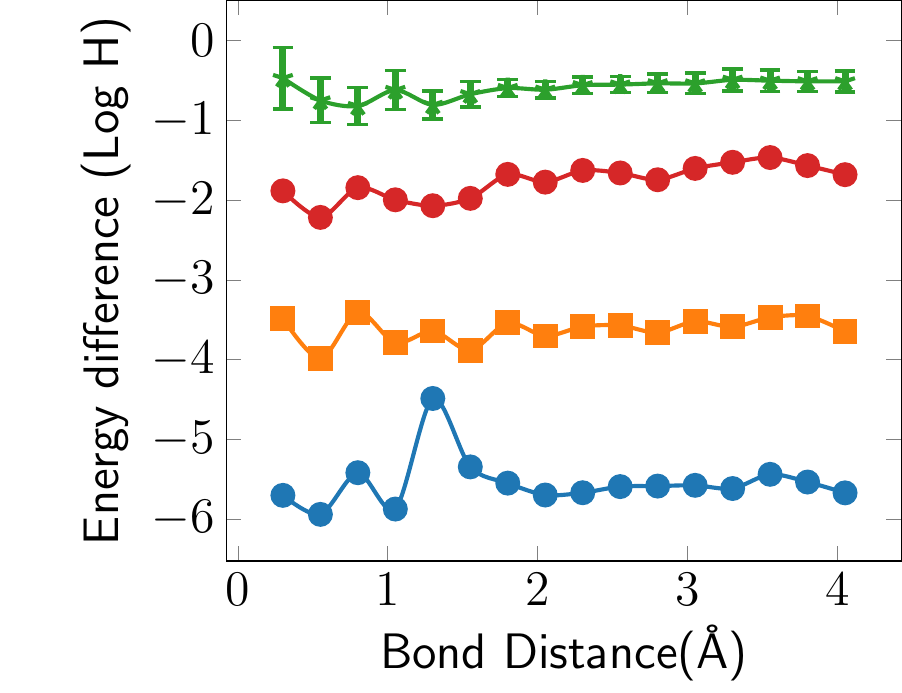}
    \caption{Energy error for the dissociation of H$_2$.
	Instead of using the exact, optimized parameters, we use parameters which
      have added Gaussian noise uncertainty. For each optimal parameter, we
      randomly sample 100 points each from a Gaussian distribution centered on
      its optimal setting and with a standard deviation of
      $\sigma=\{10^{-4},10^{-3},10^{-2},10^{-1} \}$ which correspond to the
      blue, orange, red and green curves respectively. The resulting energies
      for these 100 trials are then averaged. Error bars are given for each
      point but are smaller than some markers. 
}
\label{fig:noise1}
\end{figure}

\subsection{Noise analysis: Imprecise control}
All the analysis thus far has assumed
perfect control over
the driving parameters used to generate
approximate ground states
of chemical Hamiltonians. However, there are always
uncertainties when attempting to implement the optimal driving
parameters. In Fig.~\ref{fig:noise1} we simulate the effect of
imprecisely implementing these parameters.
For each set of optimal drive
parameters found by the {\tt ctrl-VQE} protocol,
we produce 100 samples of noisy parameters taken from Gaussian distributions
that are centered about the optimal parameter values: 
\begin{equation}
    \theta_n\leftarrow\mathrm{exp}\left(-\frac{(\theta-\theta_o)^2}{2\sigma^2}\right),
\end{equation}
where $\theta_n$ is a noisy version of the optimal parameter $\theta_o$.
These in turn are used to produce 100 noisy energy samples.
We
then average these 100 samples and find the resulting energy difference
relative to
the target ground state energy. With this applied noise model, we can see that
even imprecise settings can still be used to achieve accuracies below a
$10^{-4}$ energy error. As shown in the red and green curves in
Fig.~\ref{fig:noise1}, only when the noise becomes higher than $\sigma = 0.01$ (where errors reach a magnitude of $10~\mbox{MHz} $
or $1~\mbox{ns}$) do the energy differences become significant enough to
affect the performance of {\tt ctrl-VQE}.
Since these errors are of the
same order as the parameters themselves, they do not hinder
the realistic implementation of {\tt ctrl-VQE}.
We do not include any explicit noise due
to finite decoherence ($T_1$) or dephasing ($T_2^*$) since the pulses we find
are many orders of magnitude shorter than the typical time scales for these
effects.\cite{kochChargeinsensitiveQubitDesign2007}
To confirm this, we ran simulations using 
a Lindblad master equation that includes these
effects. Our objective function did not change significantly  due to our short time scales.

\subsection{Comparison with Circuit Compilation Techniques}\label{sec:circuit_comp}
Although several ans\"atze have been proposed to achieve shorter circuits,
	even the most compact approaches involve too many gates to implement
        on current hardware for all but the smallest molecules. 
In order to reduce the time spent during the state-preparation stage,
	and thus the coherence time demands on the hardware,
	circuit compilation techniques have been designed to take a given
        quantum circuit and execute it either with fewer 
	gates or by reoptimizing groups of gates for faster execution.

%
To execute the gate-based VQE (described in Section \ref{sec:vqe}) experimentally,
the gates in a circuit are compiled to sequences of analog control pulses
	using a look-up table that maps each elementary gate
	to the associated predefined analog pulse.
The sequence of control pulses corresponding to the  elementary gates
in the quantum circuit
are then simply collected and appropriately scheduled to be executed
on the hardware.
As such
the compilation is essentially instantaneous, making the gate-based
compilation technique well suited for VQE algorithms where numerous iterations are
performed.

From a compactness perspective however, this gate-based compilation is far from ideal,
	resulting in total pulse durations which are much longer than what
        might be obtained with optimized compilation techniques. 
As is obvious from the one-to-one translation of gates to pulses,
	the overall circuit structure is typically not considered in gate-based compilations.
Thus one may naturally be inclined to seek an optimal pulse sequence for the entire circuit.
This has motivated  compilation algorithms where control pulses are optimized
for the target circuit, using numerical optimal control techniques such as 
gradient ascent pulse engineering (GRAPE)\cite{grape} for partially optimal
compilation.\cite{Gokhale_2019} 
However, because the GRAPE algorithm itself is highly non-trivial,
	the compilation latency for each iteration is of critical concern.
Two of the GRAPE-based techniques are briefly described below; see Ref. \citenum{Gokhale_2019}
for a detailed discussion.

The GRAPE compilation technique employs an optimal control routine
which compiles to machine-level sequences of analog pulses for
a target quantum circuit. This is achieved by manipulating the sets
of time-discrete control fields that are executed on the quantum system.
The control fields in the optimal routine are updated using gradients
(see Ref. \citenum{openGrape} for use of analytical gradients) of the cost function with
respect to the control fields.
GRAPE compilation achieves up to 5-fold speedups compared to standard gate-based compilation.
However, such speedup comes with a substantial computational cost.
This amounts to long compilation latency, making this approach impractical for
VQE algorithms in which multiple iterations of circuit-parameter optimizations
are performed.
The GRAPE-based compilation also suffers
from limitations in the circuit sizes it can handle \cite{bootstrap_quantumcontrol,parallelOptimalControl,AccQOC}.

On the other hand, partial compilation techniques achieve significant pulse
speedups by leveraging the fast compilation of standard gate-based techniques
together with the pulse speedups of GRAPE-based compilations.
Two flavors of such an approach were reported in Ref. \citenum{Gokhale_2019}.
Both divide the whole circuit into blocks of subcircuits.
In the so-called strict partial compilation, the structure of quantum circuits
used in quantum variational algorithms are exploited to
only perform GRAPE-based compilation on fixed subcircuits that are independent
of the circuit parametrization. The optimal pulses using the GRAPE compilation
techniques for the fixed blocks are pre-computed and simply concatenated
with the control pulses from the gate-based compilations for the remaining
blocks of the circuit. Thus, the compilation speed is comparable to the gate-based
compilations in each iteration, but this method also takes advantage of pulse speedups
from GRAPE-based compilations. As one may expect, the pulse speedups heavily
depend on the circuit depth of the fixed blocks.

In the other flavor, flexible partial
compilation, circuits are blocked into subcircuits, each with a single
variational parameter, ensuring the subcircuit blocks have low depth.
Hyperparameter optimizations are performed for each subcircuit, and they
are utilized to find optimal pulses for the circuit blocks. It is noteworthy
to mention that in the flexible partial compilation, compilation latency is
reduced significantly (around 80x, see Ref. \citenum{Gokhale_2019}) by tuning the hyperparameters of
the circuit blocks to speed up the convergence of optimal control in GRAPE.
Flexible partial compilations achieve significant pulse speedups as
compared to strict partial compilations.
However, in spite of the
high pulse speedups, the flexible partial compilation technique still diverges
from the extremely fast compilation time of the standard gate-based methods.
Thus, the algorithm still suffers from compilation latency, albeit to a lesser degree
compared to GRAPE-based compilation.

Although both of these GRAPE-based compilation techniques and {\tt ctrl-VQE} share a direct pulse-shaping step,
	they fundamentally differ because {\tt ctrl-VQE} is not a compiler.
In contrast to full or partial compiling, we
make no reference to a circuit or to implementing any specific unitary.
Of course, the control pulses we find do implement {\it{some}} equivalent circuit
or unitary (and we analyze these in Sec. \ref{sec:decompile}), but we have no need to ever characterize the unitary.
In fact, because {\tt ctrl-VQE} only targets a single state,
a large family of unitaries (defined by the behavior in the spaces orthogonal to the initial and final states)
exist which minimize our objective function.
As such, many possible solutions exist, with no immediate preference given to any one.

\begin{figure*}[!htb] 
  \centering
  \includegraphics[width=0.90\linewidth]{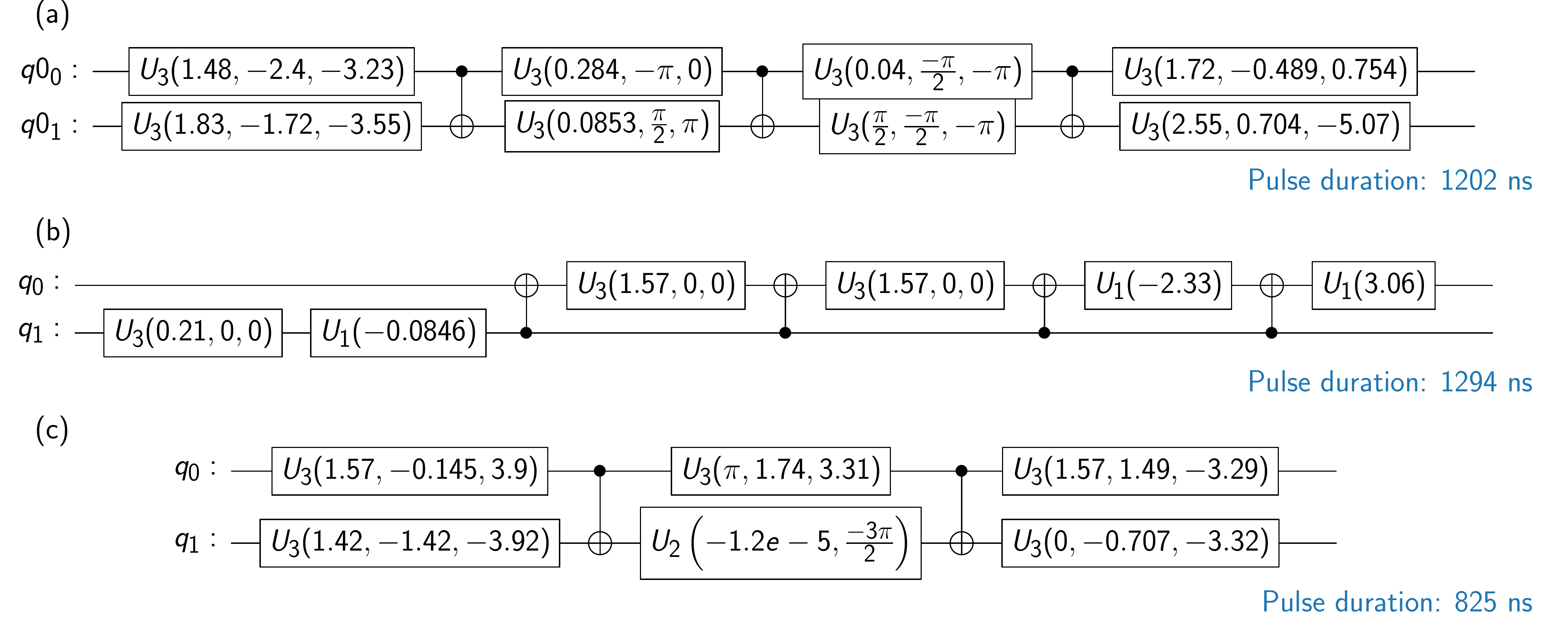}
  \caption{Illustration of circuits constructed using the unitary and state vectors from ctrl-VQE. 
	(a) Circuit corresponding to unitary obtained from a KAK decomposition. 
    	(b) An arbitary circuit corresponding to the state vector. 
	(c) Transpiled version of circuit (b).}
  \label{fig:circuit}
\end{figure*}

\subsection{Decompiled  control pulses}\label{sec:decompile}
In the previous sections, the efficiency of {\tt ctrl-VQE} 
using optimized control pulses at the device level was demonstrated. The short
pulse durations imply that applications to larger molecules
might have even more significant speedup since the number of CNOTs in
most VQE ans\"atze increases quickly.

Although {\tt ctrl-VQE} is performed using no state preparation circuit,
	the unitary created by the time dynamics of the applied pulse {\it can} be decomposed into gates,
	allowing one to analyze the circuit.
This is essentially running a compiler in reverse, or ``decompiling'' the pulse to yield a state preparation circuit.
With this decompiled circuit at hand, we can evaluate the time it would take to execute the optimized pulse as a traditional circuit.
By comparing this time to that of the pulse duration, one has a clean benchmark for quantifying the overhead associated with gate-based state preparation.
This decompilation can be done in two ways.
In the first approach, we simply compute the matrix representation of the evolution operator generated by the optimized {\tt ctrl-VQE} pulse.
For the two-qubit case in focus here, a quantum circuit is then constructed using the KAK decomposition technique.
For a detailed description of the technique, we
refer the reader to Ref. \citenum{Williams_2010}.

In the second approach, an arbitrary circuit
corresponding to the state vector from {\tt ctrl-VQE} is constructed and then
transpiled to obtain a shorter circuit depth.
In this way we obtained gate durations on
IBMQ hardware (mock Johannesburg device available in Qiskit software).
The KAK decomposition, arbitrary circuit construction,
transpilation and the mock circuit compilations were performed using Qiskit software \cite{Qiskit}.
The quantum circuits are illustrated in Figure \ref{fig:circuit}.
The corresponding pulse durations are 1202 ns for the circuit obtained using
the KAK decomposition and 825 ns for the circuit obtained from
transpilation. The state vector used in this study was for the H$_2$ molecule
at 0.75 \AA{}, and the corresponding pulse duration was 9 ns with {\tt ctrl-VQE}.
This clearly demonstrates that state preparation circuits can be much deeper
than is necessary.

\begin{figure} 
  \centering
  \includegraphics[width=0.75\linewidth]{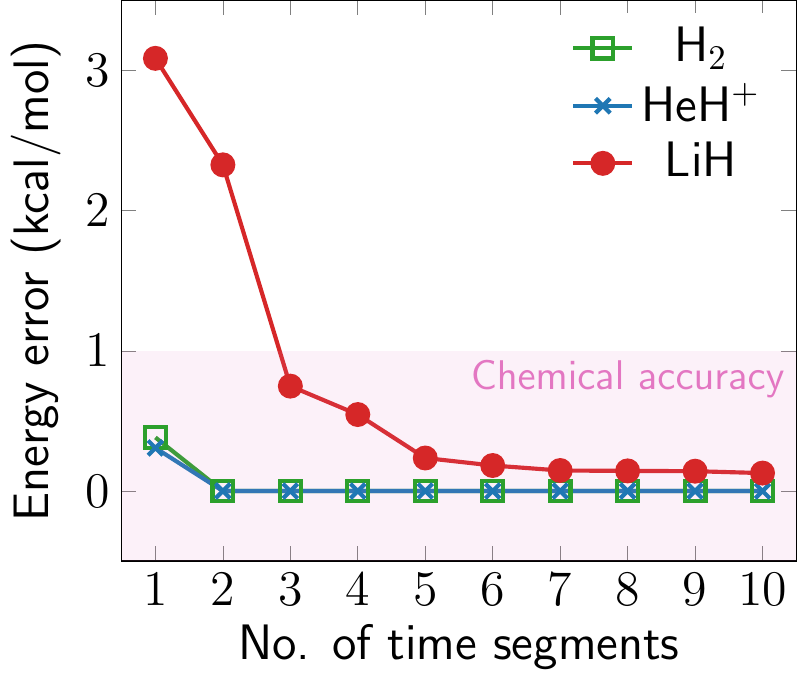}
  \caption{Convergence of pulse parameterization: Energy difference of
    {\tt ctrl-VQE} relative to FCI with increasing number of time-segments in the
    square pulse used.
        }\label{fig:adapt_pulse}
\end{figure}

\begin{table}[] 
    \centering
    \begin{tabular}{l@{\hspace{6mm}}cl}
	    \hline\hline
	    Method		&	Energy		&	Error\\
	    \hline
	    Hartree-Fock	&	-7.8633576 & 0.0177\\
	    ctrl-VQE		&	-7.8806399 & 0.0004\\
	    FCI			&	-7.8810157 & --\\
	    \hline
      Pulse duration &  40  ns     	\\ 
      Leakage        &  0.59  \%   	\\ 
      Overlap        & 99.38  \%   	\\ 
      \hline\hline
    \end{tabular}
    \caption{Performance of {\tt ctrl-VQE} in computing the
      ground state
      energy of the LiH molecule (Li-H R = 1.5 \AA).
      The energy error relative to FCI, the leakage to higher energy states,
      the total pulse duration, and the overlap with the exact FCI state are
	shown. The data correspond to the pulse with five time segments from Fig. \ref{fig:adapt_pulse}.}
    \label{table:lih} 
\end{table}

\begin{figure*}[!htb] 
  \centering
  \includegraphics[width=0.7\linewidth]{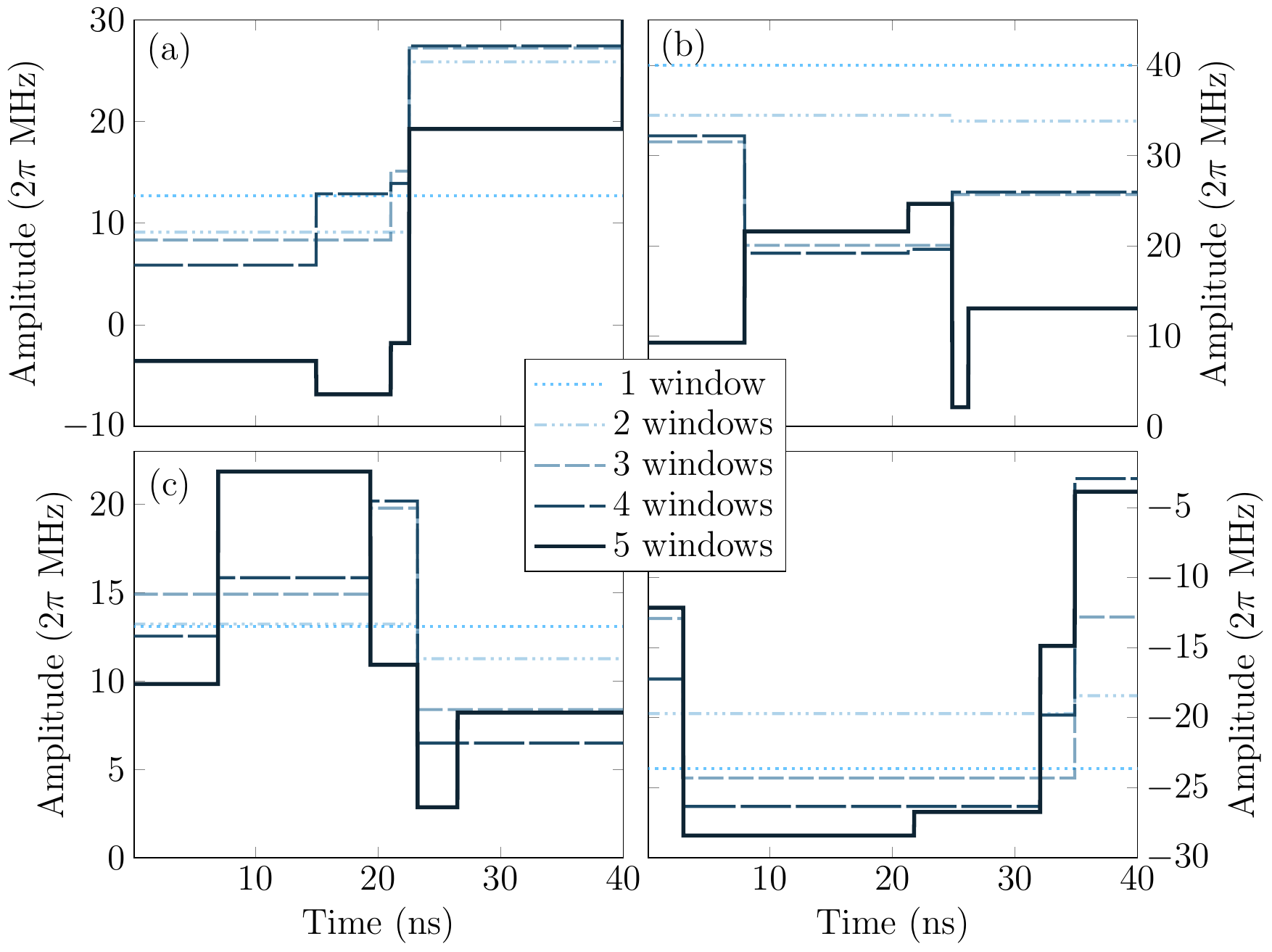}
  \caption{Adaptive increase of pulse windows for LiH molecule: Square pulse
    shapes obtained at each adaptive step, starting from one time window and
    adaptively dividing the time window up to five-segments. The accuracy
    increases with each division of the pulse. Pulse shapes are shown for the pulses on
    (a) qubit 1, (b) qubit 2, (c) qubit 3, (d) qubit 4.
        }\label{fig:adapt_pulse_shape}
\end{figure*}

{
\subsection{Adaptive update of pulse parameterization}\label{sec:adapt}
The manner in which a  pulse is parameterized (in particular, the number of variational parameters) will significantly impact experimental performance.
An over-parameterization of the pulse leads to difficulties in experimental optimization. 
In response, we seek a means to limit the number of parameters. In this section, we describe a
scheme to avoid over-parameterizations and arrive at the minimal number of
pulse parameters to achieve a target accuracy.

Unlike in the previous sections where we chose a fixed number of time segments, 
in this section we propose an adaptive algorithm which slowly grows the number of segments (and thus parameters).
We begin with a single time-segment square pulse (constant amplitude throughout the time evolution), the pulse is then
iteratively sliced at random intervals such that the number of time segments
systematically increases.
This adaptive scheme is outlined below:

\begin{enumerate}
\item Initialize a parameterized square pulse with $n=1$ time segment and
  perform the variational pulse optimization.
\item Divide the pulse obtained from the previous step at a randomly choosen
  time. 
  The square pulse now has $n=2$ time segments.
\item Perform pulse optimization using the pulse shape from the previous step
  as an initial guess.
\item Divide the largest time segment in the square pulse obtained from
  the previous step into two at a randomly chosen time. This increases the
  number of segments from $n$ to $n+1$.
\item Perform pulse optimization using the pulse shape from the previous step
  as an initial guess.
\item Repeat Steps 4 and 5 until the energy obtained is sufficiently converged.
\end{enumerate}

Note that drive frequencies for the pulses are also optimized in the above
procedure. Irrespective of the pulse shape used, only a single drive frequency
is used for each qubit. The switching times for the square pulses with
more than one time segment are not optimized and
remain fixed in Steps 3 and 5.
For $N$ transmons with $n$ pulse time segments each, the total number of
parameters to optimize is $N(n+1)$. In the following simulations, we constrain
the amplitudes and drive frequencies within the ranges $\pm 40$ MHz and
$\omega_k\pm3\pi$ GHz, respectively.


Following the above strategy of adaptively increasing the number of time
segments in the square pulses, we obtained optimal pulses for H$_2$, HeH$^+$
and LiH with total pulse durations of 9, 9 and 40 ns
respectively. Fig. \ref{fig:adapt_pulse} plots the energy difference relative to FCI
with increasing number of time segments in the optimal pulse. For H$_2$ and
HeH$^+$, chemical accuracy is obtained with a single-time-segment
square pulse on each qubit, while for LiH, three time segments are required to
achieve chemical accuracy. The molecular energy converges for square
pulses with two time segments for H$_2$ and HeH$^+$ and with five time
segments for LiH molecules.
The error in ground state energy of LiH molecule (Li-H R=1.5 \AA) computed with
{\tt ctrl-VQE} along with pulse duration, leakage and overlap with the exact
FCI state is tabulated in Table \ref{table:lih}.
To illustrate the sequential pulse refinement from iteration to iteration,
the square pulse shapes at each adaptive step for
LiH are shown in Fig. \ref{fig:adapt_pulse_shape}.

In the case of LiH, we see that the accuracy is not of the same order as in
the other two diatomic molecules. To further examine this, we perform over
100 pulse optimizations on the square pulse with 10 time segments obtained from
the adaptive strategy by retaining only the switching times and starting with
random amplitudes and drive frequencies.
We use a large number of initial guesses to reduce the likelihood of getting
trapped in a local minimum.
  We find that the highest accuracy is achieved when we use
the pulse obtained from the
adaptive strategy above. This suggests that we are nearing the accuracy limit
of LiH using a 40 ns pulse.

While the above adaptive strategy aims at practical experimental
implementation, the cost of classical simulation remains the same irrespective
of the number of time-segments when analytical gradients for the pulse amplitudes are used.
The cost of evaluating an analytical gradient is only about 2.5 times
the cost of an energy evaluation. Here, the analytical gradient is obtained at
each trotter step of time evolution and adapted for the pulse shaped
considered. Details along with the derivation are provided in the SI.
}

\subsection{Comparison to gate-based ans\"atze}
Now we directly compare the results of our pulse-based technique with
gate-based variational ans\"atze. 
We use calibration data from an IBMQ device
(mock Johannesburg device available in Qiskit software\footnote{\label{ll}The mock
  Johannesburg device was taken from Qiskit Terra version 0.14.2}) to compute
the
duration of the circuits used to prepare trial wavefunctions. We consider the
RY and UCCSD ans\"atze, which are capable of producing the exact ground state.

For both H$_2$ and HeH$^+$, the RY ansatz requires 1 CNOT, and the UCCSD requires 2 CNOTs.
The RY ansatz has 4 parameters, and the UCCSD ansatz has 3 parameters.
The total pulse time of the RY ansatz is 519 ns, and the UCCSD ansatz is 825 ns. 
For the LiH molecule, the RY and UCCSD ans\"atz require 9 and 245 CNOTs
respectively. The RY ansatz has 16 parameters whereas the UCCSD ansatz has 8
parameters. The total pulse duration of the RY ansatz is 3485 ns,
and that of the UCCSD ansatz is 82169 ns for LiH.
In each case, the time to execute the circuit is significantly longer than the
time required to apply the pulses using { {\tt ctrl-VQE},
which is 9 ns for H$_2$ and HeH$^+$, and 40 ns for LiH.
The $T_1$ and $T_2$ times of the IBMQ device used range from 45150 to 115217
ns (average 71262 ns) and from 48173 to 107025 ns (average 74490 ns)
respectively. This means that the pulse duration for LiH using the UCCSD
ansatz is already reaching the limit set by decoherence times of the device,
while the pulse duration obtained with {\tt ctrl-VQE} is well within the
decoherence time.}

We note here that this comparison needs to be approached with a bit of
caution, as our simulated device has slightly different parameters than the
IBM-Q devices used for the circuit timings.
However, we do not expect that our results would change significantly if we had
access to the full set of exact parameters of the device
(including anharmonicities, bus frequencies, and couplings), because
previous works using the present device parameter
regime\cite{PhysRevB.91.161405, PhysRevB.101.054508} have demonstrated
universal gate sets with single and two-qubit operations commensurate with
those of the IBM-Q devices.
As such, we expect the {\tt ctrl-VQE} pulse times reported in this study to be
close to
what one would get by directly running the calculations on IBM-Q devices. 

\section{Summary and Future Outlook}\label{sec:summary}
We have presented a new variational quantum algorithm which is fundamentally different from the
existing quantum algorithms for molecular simulation. 
The quantum circuit used for state preparation in standard
variational algorithms is entirely replaced by a hardware-level control
routine with optimized pulse shapes to directly drive the qubit
system. The efficacy of the presented method is numerically demonstrated by
modeling the bond dissociation of two diatomic molecules, H$_2$ and
HeH$^+$. The maximum error from the exact FCI energy is well within chemical accuracy (0.02 kcal/mol)
for both molecular systems throughout the bond dissociation.
{\tt ctrl-VQE} captures the important electron correlation effects involved in the bond dissociation of
the two molecular systems, as reflected in the pulse durations along the
bond dissociation.
{
To demonstrate the application of {\tt ctrl-VQE} to a larger molecular system, we have
also computed the energy of the LiH molecule at a single bond distance. An
adaptive scheme which slowly increases the variational flexibility of the pulse
parameterization was used to determine the minimal number of pulse parameters
required to acheive a target accuracy, ultimately simplifying experimental
implementation and avoiding over-parameterization. The results demonstrate
that pulse shapes with relatively modest numbers of parameters achieve
convergence in the energy even as the problem size increased.}
The approach yields significant state-preparation speedups for VQE as compared to standard gate-based ans\"atze.
The short pulse durations minimize losses
due to decoherence and dephasing, which is a step toward enabling more accurate VQE simulations on larger, strongly correlated systems.
{\tt ctrl-VQE} can be viewed as a lowest-level quantum variational algorithm.

Because {\tt ctrl-VQE} operates directly on the hardware,
classical simulations of the algorithm
are even more computationally expensive than typical VQE emulations,
	as one needs to not only solve the time-dependent Schr\"odinger equation, but also optimize the driving Hamiltonian.
As such, the systems we have been able to study so far are very small.
In future work, we will develop an improved implementation to study the behavior of larger systems
and with more sophisticated constraints on the pulse shape ($\Omega_n(t)$).
Improved software will also make it possible to numerically study the impact that device controllability limitations\cite{ballSoftwareToolsQuantum2020}
have on the amount or type of electron correlation able to be described.


\section*{Competing Interest Statement}
The Authors declare no Competing Financial or Non-Financial Interests.

\section*{Data Availability}
The data for the numerical simulations is available upon reasonable request.

\section*{Code Availability}
The code for the numerical simulations is available at \url{https://github.com/mayhallgroup/ctrlq}.

\section*{Author Contributions}
N.J.M., S.E.E, E.B., and D.P.P. conceived the project. O.R.M. wrote the python/C++ code and helped design the adaptive pulse parameterization algorithm. B.T.G. wrote mathematica code. B.T.G. and O.R.M performed numerical experiments. All authors contributed to the conceptual analysis of the results and to ideas that refined the algorithm. All authors wrote the paper.

\section{Acknowledgements}
N.J.M., S.E.E., E.B., and D.P.P are grateful for financial support provided by the U.S. Department
of Energy (Award No. DE-SC0019199). 
S.E.E acknowledges support provided by the U.S. Department of Energy (Award No. DE-SC0019318). 
This work was done in part while N.J.M. was visiting the Simons Institute for the Theory of Computing.
The authors thanks the Advanced Research Computing at Virginia Tech for the computational infrastructure.


\begin{thebibliography}{69}
\expandafter\ifx\csname natexlab\endcsname\relax\def\natexlab#1{#1}\fi
\expandafter\ifx\csname bibnamefont\endcsname\relax
  \def\bibnamefont#1{#1}\fi
\expandafter\ifx\csname bibfnamefont\endcsname\relax
  \def\bibfnamefont#1{#1}\fi
\expandafter\ifx\csname citenamefont\endcsname\relax
  \def\citenamefont#1{#1}\fi
\expandafter\ifx\csname url\endcsname\relax
  \def\url#1{\texttt{#1}}\fi
\expandafter\ifx\csname urlprefix\endcsname\relax\def\urlprefix{URL }\fi
\providecommand{\bibinfo}[2]{#2}
\providecommand{\eprint}[2][]{\url{#2}}

\bibitem[{\citenamefont{White}(1992)}]{whiteDensityMatrixFormulation1992}
\bibinfo{author}{\bibfnamefont{S.~R.} \bibnamefont{White}},
  \bibinfo{journal}{Physical Review Letters} \textbf{\bibinfo{volume}{69}},
  \bibinfo{pages}{2863} (\bibinfo{year}{1992}).

\bibitem[{\citenamefont{Chan and
  {Head-Gordon}}(2002)}]{chanHighlyCorrelatedCalculations2002}
\bibinfo{author}{\bibfnamefont{G.~K.-L.} \bibnamefont{Chan}} \bibnamefont{and}
  \bibinfo{author}{\bibfnamefont{M.}~\bibnamefont{{Head-Gordon}}},
  \bibinfo{journal}{J. Chem. Phys.} \textbf{\bibinfo{volume}{116}},
  \bibinfo{pages}{4462} (\bibinfo{year}{2002}), ISSN \bibinfo{issn}{0021-9606}.

\bibitem[{\citenamefont{Schollw{\"o}ck}(2011)}]{schollwockDensitymatrixRenormalizationGroup2011}
\bibinfo{author}{\bibfnamefont{U.}~\bibnamefont{Schollw{\"o}ck}},
  \bibinfo{journal}{Annals of Physics} \textbf{\bibinfo{volume}{326}},
  \bibinfo{pages}{96} (\bibinfo{year}{2011}).

\bibitem[{\citenamefont{Huron et~al.}(1973)\citenamefont{Huron, Malrieu, and
  Rancurel}}]{Huron1973}
\bibinfo{author}{\bibfnamefont{B.}~\bibnamefont{Huron}},
  \bibinfo{author}{\bibfnamefont{J.~P.} \bibnamefont{Malrieu}},
  \bibnamefont{and} \bibinfo{author}{\bibfnamefont{P.}~\bibnamefont{Rancurel}},
  \bibinfo{journal}{J. Chem. Phys.} \textbf{\bibinfo{volume}{58}},
  \bibinfo{pages}{5745} (\bibinfo{year}{1973}), ISSN \bibinfo{issn}{00219606},
  \urlprefix\url{http://link.aip.org/link/?JCPSA6/58/5745/1}.

\bibitem[{\citenamefont{Bender and Davidson}(1969)}]{Bender1969}
\bibinfo{author}{\bibfnamefont{C.~F.} \bibnamefont{Bender}} \bibnamefont{and}
  \bibinfo{author}{\bibfnamefont{E.~R.} \bibnamefont{Davidson}},
  \bibinfo{journal}{Phys. Rev.} \textbf{\bibinfo{volume}{183}},
  \bibinfo{pages}{23} (\bibinfo{year}{1969}),
  \urlprefix\url{https://link.aps.org/doi/10.1103/PhysRev.183.23}.

\bibitem[{\citenamefont{Buenker}(1968)}]{Buenker1968}
\bibinfo{author}{\bibfnamefont{R.~J.} \bibnamefont{Buenker}},
  \bibinfo{journal}{J. Chem. Phys.} \textbf{\bibinfo{volume}{49}},
  \bibinfo{pages}{5381} (\bibinfo{year}{1968}), ISSN \bibinfo{issn}{00219606}.

\bibitem[{\citenamefont{Evangelisti et~al.}(1983)\citenamefont{Evangelisti,
  Daudey, and Malrieu}}]{evangelistiConvergenceImprovedCIPSI1983}
\bibinfo{author}{\bibfnamefont{S.}~\bibnamefont{Evangelisti}},
  \bibinfo{author}{\bibfnamefont{J.-P.} \bibnamefont{Daudey}},
  \bibnamefont{and} \bibinfo{author}{\bibfnamefont{J.-P.}
  \bibnamefont{Malrieu}}, \bibinfo{journal}{Chemical Physics}
  \textbf{\bibinfo{volume}{75}}, \bibinfo{pages}{91} (\bibinfo{year}{1983}),
  ISSN \bibinfo{issn}{0301-0104}.

\bibitem[{\citenamefont{Tubman et~al.}(2016)\citenamefont{Tubman, Lee,
  Takeshita, {Head-Gordon}, and
  Whaley}}]{tubmanDeterministicAlternativeFull2016}
\bibinfo{author}{\bibfnamefont{N.~M.} \bibnamefont{Tubman}},
  \bibinfo{author}{\bibfnamefont{J.}~\bibnamefont{Lee}},
  \bibinfo{author}{\bibfnamefont{T.~Y.} \bibnamefont{Takeshita}},
  \bibinfo{author}{\bibfnamefont{M.}~\bibnamefont{{Head-Gordon}}},
  \bibnamefont{and} \bibinfo{author}{\bibfnamefont{K.~B.}
  \bibnamefont{Whaley}}, \bibinfo{journal}{J. Chem. Phys.}
  \textbf{\bibinfo{volume}{145}}, \bibinfo{pages}{044112}
  (\bibinfo{year}{2016}), ISSN \bibinfo{issn}{0021-9606}.

\bibitem[{\citenamefont{Schriber and
  Evangelista}(2016)}]{schriberCommunicationAdaptiveConfiguration2016}
\bibinfo{author}{\bibfnamefont{J.~B.} \bibnamefont{Schriber}} \bibnamefont{and}
  \bibinfo{author}{\bibfnamefont{F.~A.} \bibnamefont{Evangelista}},
  \bibinfo{journal}{The Journal of Chemical Physics}
  \textbf{\bibinfo{volume}{144}}, \bibinfo{pages}{161106}
  (\bibinfo{year}{2016}).

\bibitem[{\citenamefont{Holmes et~al.}(2016)\citenamefont{Holmes, Tubman, and
  Umrigar}}]{holmesHeatBathConfigurationInteraction2016}
\bibinfo{author}{\bibfnamefont{A.~A.} \bibnamefont{Holmes}},
  \bibinfo{author}{\bibfnamefont{N.~M.} \bibnamefont{Tubman}},
  \bibnamefont{and} \bibinfo{author}{\bibfnamefont{C.~J.}
  \bibnamefont{Umrigar}}, \bibinfo{journal}{Journal of Chemical Theory and
  Computation} \textbf{\bibinfo{volume}{12}}, \bibinfo{pages}{3674}
  (\bibinfo{year}{2016}).

\bibitem[{\citenamefont{Levine et~al.}(2020)\citenamefont{Levine, Hait, Tubman,
  Lehtola, Whaley, and {Head-Gordon}}}]{levineCASSCFExtremelyLarge2020}
\bibinfo{author}{\bibfnamefont{D.~S.} \bibnamefont{Levine}},
  \bibinfo{author}{\bibfnamefont{D.}~\bibnamefont{Hait}},
  \bibinfo{author}{\bibfnamefont{N.~M.} \bibnamefont{Tubman}},
  \bibinfo{author}{\bibfnamefont{S.}~\bibnamefont{Lehtola}},
  \bibinfo{author}{\bibfnamefont{K.~B.} \bibnamefont{Whaley}},
  \bibnamefont{and}
  \bibinfo{author}{\bibfnamefont{M.}~\bibnamefont{{Head-Gordon}}},
  \bibinfo{journal}{J. Chem. Theory Comput.} \textbf{\bibinfo{volume}{16}},
  \bibinfo{pages}{2340} (\bibinfo{year}{2020}), ISSN \bibinfo{issn}{1549-9618}.

\bibitem[{\citenamefont{Caffarel et~al.}(2016)\citenamefont{Caffarel,
  Applencourt, Giner, and Scemama}}]{caffarelUsingCIPSINodes2016}
\bibinfo{author}{\bibfnamefont{M.}~\bibnamefont{Caffarel}},
  \bibinfo{author}{\bibfnamefont{T.}~\bibnamefont{Applencourt}},
  \bibinfo{author}{\bibfnamefont{E.}~\bibnamefont{Giner}}, \bibnamefont{and}
  \bibinfo{author}{\bibfnamefont{A.}~\bibnamefont{Scemama}}, in
  \emph{\bibinfo{booktitle}{Recent {{Progress}} in {{Quantum Monte Carlo}}}}
  (\bibinfo{publisher}{{American Chemical Society}}, \bibinfo{year}{2016}),
  vol. \bibinfo{volume}{1234} of \emph{\bibinfo{series}{{{ACS Symposium
  Series}}}}, chap.~\bibinfo{chapter}{2}, pp. \bibinfo{pages}{15--46}, ISBN
  \bibinfo{isbn}{978-0-8412-3179-5}.

\bibitem[{\citenamefont{Abraham and
  Mayhall}(2020)}]{abrahamSelectedConfigurationInteraction2020c}
\bibinfo{author}{\bibfnamefont{V.}~\bibnamefont{Abraham}} \bibnamefont{and}
  \bibinfo{author}{\bibfnamefont{N.~J.} \bibnamefont{Mayhall}},
  \bibinfo{journal}{J. Chem. Theory Comput.} \textbf{\bibinfo{volume}{16}},
  \bibinfo{pages}{6098} (\bibinfo{year}{2020}), ISSN \bibinfo{issn}{1549-9618}.

\bibitem[{\citenamefont{Peruzzo et~al.}(2014)\citenamefont{Peruzzo, McClean,
  Shadbolt, Yung, Zhou, Love, {Aspuru-Guzik}, and
  O'Brien}}]{peruzzoVariationalEigenvalueSolver2014}
\bibinfo{author}{\bibfnamefont{A.}~\bibnamefont{Peruzzo}},
  \bibinfo{author}{\bibfnamefont{J.}~\bibnamefont{McClean}},
  \bibinfo{author}{\bibfnamefont{P.}~\bibnamefont{Shadbolt}},
  \bibinfo{author}{\bibfnamefont{M.-H.} \bibnamefont{Yung}},
  \bibinfo{author}{\bibfnamefont{X.-Q.} \bibnamefont{Zhou}},
  \bibinfo{author}{\bibfnamefont{P.~J.} \bibnamefont{Love}},
  \bibinfo{author}{\bibfnamefont{A.}~\bibnamefont{{Aspuru-Guzik}}},
  \bibnamefont{and} \bibinfo{author}{\bibfnamefont{J.~L.}
  \bibnamefont{O'Brien}}, \bibinfo{journal}{Nature communications}
  \textbf{\bibinfo{volume}{5}}, \bibinfo{pages}{4213} (\bibinfo{year}{2014}),
  ISSN \bibinfo{issn}{2041-1723 (Electronic) 2041-1723 (Linking)}.

\bibitem[{\citenamefont{Preskill}(2018)}]{preskillQuantumComputingNISQ2018}
\bibinfo{author}{\bibfnamefont{J.}~\bibnamefont{Preskill}},
  \bibinfo{journal}{Quantum} \textbf{\bibinfo{volume}{2}}, \bibinfo{pages}{79}
  (\bibinfo{year}{2018}), ISSN \bibinfo{issn}{2521-327X}.

\bibitem[{\citenamefont{Sharma et~al.}(2020)\citenamefont{Sharma, Khatri,
  Cerezo, and Coles}}]{Sharma_2020}
\bibinfo{author}{\bibfnamefont{K.}~\bibnamefont{Sharma}},
  \bibinfo{author}{\bibfnamefont{S.}~\bibnamefont{Khatri}},
  \bibinfo{author}{\bibfnamefont{M.}~\bibnamefont{Cerezo}}, \bibnamefont{and}
  \bibinfo{author}{\bibfnamefont{P.~J.} \bibnamefont{Coles}},
  \bibinfo{journal}{New Journal of Physics} \textbf{\bibinfo{volume}{22}},
  \bibinfo{pages}{043006} (\bibinfo{year}{2020}),
  \urlprefix\url{https://doi.org/10.1088%2F1367-2630%2Fab784c}.

\bibitem[{\citenamefont{Kandala et~al.}(2017)\citenamefont{Kandala, Mezzacapo,
  Temme, Takita, Brink, Chow, and Gambetta}}]{Kandala_2017}
\bibinfo{author}{\bibfnamefont{A.}~\bibnamefont{Kandala}},
  \bibinfo{author}{\bibfnamefont{A.}~\bibnamefont{Mezzacapo}},
  \bibinfo{author}{\bibfnamefont{K.}~\bibnamefont{Temme}},
  \bibinfo{author}{\bibfnamefont{M.}~\bibnamefont{Takita}},
  \bibinfo{author}{\bibfnamefont{M.}~\bibnamefont{Brink}},
  \bibinfo{author}{\bibfnamefont{J.~M.} \bibnamefont{Chow}}, \bibnamefont{and}
  \bibinfo{author}{\bibfnamefont{J.~M.} \bibnamefont{Gambetta}},
  \bibinfo{journal}{Nature} \textbf{\bibinfo{volume}{549}},
  \bibinfo{pages}{242} (\bibinfo{year}{2017}),
  \urlprefix\url{https://doi.org/10.1038/nature23879}.

\bibitem[{\citenamefont{Huggins
  et~al.}(2019{\natexlab{a}})\citenamefont{Huggins, Lee, Baek, O'Gorman, and
  Whaley}}]{hugginsNonOrthogonalVariationalQuantum2019a}
\bibinfo{author}{\bibfnamefont{W.~J.} \bibnamefont{Huggins}},
  \bibinfo{author}{\bibfnamefont{J.}~\bibnamefont{Lee}},
  \bibinfo{author}{\bibfnamefont{U.}~\bibnamefont{Baek}},
  \bibinfo{author}{\bibfnamefont{B.}~\bibnamefont{O'Gorman}}, \bibnamefont{and}
  \bibinfo{author}{\bibfnamefont{K.~B.} \bibnamefont{Whaley}},
  \bibinfo{journal}{arXiv:1909.09114 [physics, physics:quant-ph]}
  (\bibinfo{year}{2019}{\natexlab{a}}), \eprint{1909.09114}.

\bibitem[{\citenamefont{Lee et~al.}(2019)\citenamefont{Lee, Huggins,
  {Head-Gordon}, and Whaley}}]{leeGeneralizedUnitaryCoupled2019}
\bibinfo{author}{\bibfnamefont{J.}~\bibnamefont{Lee}},
  \bibinfo{author}{\bibfnamefont{W.~J.} \bibnamefont{Huggins}},
  \bibinfo{author}{\bibfnamefont{M.}~\bibnamefont{{Head-Gordon}}},
  \bibnamefont{and} \bibinfo{author}{\bibfnamefont{K.~B.}
  \bibnamefont{Whaley}}, \bibinfo{journal}{J. Chem. Theory Comput.}
  \textbf{\bibinfo{volume}{15}}, \bibinfo{pages}{311} (\bibinfo{year}{2019}),
  ISSN \bibinfo{issn}{1549-9618}.

\bibitem[{\citenamefont{Gard et~al.}(2020)\citenamefont{Gard, Zhu, Barron,
  Mayhall, Economou, and Barnes}}]{gardEfficientSymmetrypreservingState2020}
\bibinfo{author}{\bibfnamefont{B.~T.} \bibnamefont{Gard}},
  \bibinfo{author}{\bibfnamefont{L.}~\bibnamefont{Zhu}},
  \bibinfo{author}{\bibfnamefont{G.~S.} \bibnamefont{Barron}},
  \bibinfo{author}{\bibfnamefont{N.~J.} \bibnamefont{Mayhall}},
  \bibinfo{author}{\bibfnamefont{S.~E.} \bibnamefont{Economou}},
  \bibnamefont{and} \bibinfo{author}{\bibfnamefont{E.}~\bibnamefont{Barnes}},
  \bibinfo{journal}{npj Quantum Information} \textbf{\bibinfo{volume}{6}},
  \bibinfo{pages}{1} (\bibinfo{year}{2020}), ISSN \bibinfo{issn}{2056-6387}.

\bibitem[{\citenamefont{Barron et~al.}(2020{\natexlab{a}})\citenamefont{Barron,
  Gard, Altman, Mayhall, Barnes, and Economou}}]{Barron_2020}
\bibinfo{author}{\bibfnamefont{G.~S.} \bibnamefont{Barron}},
  \bibinfo{author}{\bibfnamefont{B.~T.} \bibnamefont{Gard}},
  \bibinfo{author}{\bibfnamefont{O.~J.} \bibnamefont{Altman}},
  \bibinfo{author}{\bibfnamefont{N.~J.} \bibnamefont{Mayhall}},
  \bibinfo{author}{\bibfnamefont{E.}~\bibnamefont{Barnes}}, \bibnamefont{and}
  \bibinfo{author}{\bibfnamefont{S.~E.} \bibnamefont{Economou}},
  \bibinfo{journal}{arXiv:2003.00171 [quant-ph]}
  (\bibinfo{year}{2020}{\natexlab{a}}).

\bibitem[{\citenamefont{Ryabinkin et~al.}(2018)\citenamefont{Ryabinkin, Yen,
  Genin, and Izmaylov}}]{ryabinkinQubitCoupledclusterMethod2018}
\bibinfo{author}{\bibfnamefont{I.~G.} \bibnamefont{Ryabinkin}},
  \bibinfo{author}{\bibfnamefont{T.-C.} \bibnamefont{Yen}},
  \bibinfo{author}{\bibfnamefont{S.~N.} \bibnamefont{Genin}}, \bibnamefont{and}
  \bibinfo{author}{\bibfnamefont{A.~F.} \bibnamefont{Izmaylov}}
  (\bibinfo{year}{2018}).

\bibitem[{\citenamefont{Grimsley et~al.}(2019)\citenamefont{Grimsley, Economou,
  Barnes, and Mayhall}}]{grimsleyAdaptiveVariationalAlgorithm2019}
\bibinfo{author}{\bibfnamefont{H.~R.} \bibnamefont{Grimsley}},
  \bibinfo{author}{\bibfnamefont{S.~E.} \bibnamefont{Economou}},
  \bibinfo{author}{\bibfnamefont{E.}~\bibnamefont{Barnes}}, \bibnamefont{and}
  \bibinfo{author}{\bibfnamefont{N.~J.} \bibnamefont{Mayhall}},
  \bibinfo{journal}{Nat. Commun.} \textbf{\bibinfo{volume}{10}},
  \bibinfo{pages}{3007} (\bibinfo{year}{2019}), ISSN \bibinfo{issn}{2041-1723}.

\bibitem[{\citenamefont{Tang et~al.}(2020)\citenamefont{Tang, Shkolnikov,
  Barron, Grimsley, Mayhall, Barnes, and
  Economou}}]{tangQubitADAPTVQEAdaptiveAlgorithm2020}
\bibinfo{author}{\bibfnamefont{H.~L.} \bibnamefont{Tang}},
  \bibinfo{author}{\bibfnamefont{V.~O.} \bibnamefont{Shkolnikov}},
  \bibinfo{author}{\bibfnamefont{G.~S.} \bibnamefont{Barron}},
  \bibinfo{author}{\bibfnamefont{H.~R.} \bibnamefont{Grimsley}},
  \bibinfo{author}{\bibfnamefont{N.~J.} \bibnamefont{Mayhall}},
  \bibinfo{author}{\bibfnamefont{E.}~\bibnamefont{Barnes}}, \bibnamefont{and}
  \bibinfo{author}{\bibfnamefont{S.~E.} \bibnamefont{Economou}},
  \bibinfo{journal}{arXiv:1911.10205 [quant-ph]}  (\bibinfo{year}{2020}),
  \eprint{1911.10205}.

\bibitem[{\citenamefont{Ryabinkin et~al.}(2020)\citenamefont{Ryabinkin, Lang,
  Genin, and Izmaylov}}]{ryabinkinIterativeQubitCoupled2020}
\bibinfo{author}{\bibfnamefont{I.~G.} \bibnamefont{Ryabinkin}},
  \bibinfo{author}{\bibfnamefont{R.~A.} \bibnamefont{Lang}},
  \bibinfo{author}{\bibfnamefont{S.~N.} \bibnamefont{Genin}}, \bibnamefont{and}
  \bibinfo{author}{\bibfnamefont{A.~F.} \bibnamefont{Izmaylov}},
  \bibinfo{journal}{J. Chem. Theory Comput.} \textbf{\bibinfo{volume}{16}},
  \bibinfo{pages}{1055} (\bibinfo{year}{2020}), ISSN \bibinfo{issn}{1549-9618}.

\bibitem[{\citenamefont{Jordan}(1928)}]{jordanwigner1928}
\bibinfo{author}{\bibfnamefont{W.}~\bibnamefont{Jordan}, \bibfnamefont{P.}},
  \bibinfo{journal}{Z. Physik} \textbf{\bibinfo{volume}{47}},
  \bibinfo{pages}{631} (\bibinfo{year}{1928}),
  \urlprefix\url{https://doi.org/10.1007/BF01331938}.

\bibitem[{\citenamefont{Bravyi and Kitaev}(2002)}]{bravyi2002}
\bibinfo{author}{\bibfnamefont{S.~B.} \bibnamefont{Bravyi}} \bibnamefont{and}
  \bibinfo{author}{\bibfnamefont{A.~Y.} \bibnamefont{Kitaev}},
  \bibinfo{journal}{Ann. Phys.} \textbf{\bibinfo{volume}{298}},
  \bibinfo{pages}{210} (\bibinfo{year}{2002}),
  \urlprefix\url{https://doi.org/10.1006/aphy.2002.6254}.

\bibitem[{\citenamefont{Seeley et~al.}(2012)\citenamefont{Seeley, Richard, and
  Love}}]{parity}
\bibinfo{author}{\bibfnamefont{J.~T.} \bibnamefont{Seeley}},
  \bibinfo{author}{\bibfnamefont{M.~J.} \bibnamefont{Richard}},
  \bibnamefont{and} \bibinfo{author}{\bibfnamefont{P.~J.} \bibnamefont{Love}},
  \bibinfo{journal}{The Journal of Chemical Physics}
  \textbf{\bibinfo{volume}{137}}, \bibinfo{pages}{224109}
  (\bibinfo{year}{2012}), \eprint{https://doi.org/10.1063/1.4768229},
  \urlprefix\url{https://doi.org/10.1063/1.4768229}.

\bibitem[{\citenamefont{Bartlett and Musia\l{}}(2007)}]{uccsd_bartlett}
\bibinfo{author}{\bibfnamefont{R.~J.} \bibnamefont{Bartlett}} \bibnamefont{and}
  \bibinfo{author}{\bibfnamefont{M.}~\bibnamefont{Musia\l{}}},
  \bibinfo{journal}{Rev. Mod. Phys.} \textbf{\bibinfo{volume}{79}},
  \bibinfo{pages}{291} (\bibinfo{year}{2007}),
  \urlprefix\url{https://link.aps.org/doi/10.1103/RevModPhys.79.291}.

\bibitem[{\citenamefont{Yung et~al.}(2014)\citenamefont{Yung, Casanova,
  Mezzacapo, McClean, Lamata, and Aspuru-Guzik}}]{uccsd_vqe_alan}
\bibinfo{author}{\bibfnamefont{M.-H.} \bibnamefont{Yung}},
  \bibinfo{author}{\bibfnamefont{J.}~\bibnamefont{Casanova}},
  \bibinfo{author}{\bibfnamefont{A.}~\bibnamefont{Mezzacapo}},
  \bibinfo{author}{\bibfnamefont{J.}~\bibnamefont{McClean}},
  \bibinfo{author}{\bibfnamefont{L.}~\bibnamefont{Lamata}}, \bibnamefont{and}
  \bibinfo{author}{\bibfnamefont{E.}~\bibnamefont{Aspuru-Guzik},
  \bibfnamefont{A.and~Solano}}, \bibinfo{journal}{Scientific Reports}
  \textbf{\bibinfo{volume}{4}}, \bibinfo{pages}{3589} (\bibinfo{year}{2014}).

\bibitem[{\citenamefont{Ostaszewski et~al.}(2019)\citenamefont{Ostaszewski,
  Grant, and Benedetti}}]{ostaszewskiQuantumCircuitStructure2019}
\bibinfo{author}{\bibfnamefont{M.}~\bibnamefont{Ostaszewski}},
  \bibinfo{author}{\bibfnamefont{E.}~\bibnamefont{Grant}}, \bibnamefont{and}
  \bibinfo{author}{\bibfnamefont{M.}~\bibnamefont{Benedetti}},
  \bibinfo{journal}{arXiv:1905.09692 [quant-ph]}  (\bibinfo{year}{2019}),
  \eprint{1905.09692}.

\bibitem[{\citenamefont{Chivilikhin et~al.}(2020)\citenamefont{Chivilikhin,
  Samarin, Ulyantsev, Iorsh, Oganov, and
  Kyriienko}}]{chivilikhinMoGVQEMultiobjectiveGenetic2020}
\bibinfo{author}{\bibfnamefont{D.}~\bibnamefont{Chivilikhin}},
  \bibinfo{author}{\bibfnamefont{A.}~\bibnamefont{Samarin}},
  \bibinfo{author}{\bibfnamefont{V.}~\bibnamefont{Ulyantsev}},
  \bibinfo{author}{\bibfnamefont{I.}~\bibnamefont{Iorsh}},
  \bibinfo{author}{\bibfnamefont{A.~R.} \bibnamefont{Oganov}},
  \bibnamefont{and}
  \bibinfo{author}{\bibfnamefont{O.}~\bibnamefont{Kyriienko}},
  \bibinfo{journal}{arXiv:2007.04424 [cond-mat, physics:quant-ph]}
  (\bibinfo{year}{2020}), \eprint{2007.04424}.

\bibitem[{\citenamefont{Matsuzawa and
  Kurashige}(2020)}]{jastrow_low_depth_Yuta}
\bibinfo{author}{\bibfnamefont{Y.}~\bibnamefont{Matsuzawa}} \bibnamefont{and}
  \bibinfo{author}{\bibfnamefont{Y.}~\bibnamefont{Kurashige}},
  \bibinfo{journal}{Journal of Chemical Theory and Computation}
  \textbf{\bibinfo{volume}{16}}, \bibinfo{pages}{944} (\bibinfo{year}{2020}).

\bibitem[{\citenamefont{Huggins
  et~al.}(2019{\natexlab{b}})\citenamefont{Huggins, Lee, Baek, O'Gorman, and
  Whaley}}]{hugginsNonOrthogonalVariationalQuantum2019}
\bibinfo{author}{\bibfnamefont{W.~J.} \bibnamefont{Huggins}},
  \bibinfo{author}{\bibfnamefont{J.}~\bibnamefont{Lee}},
  \bibinfo{author}{\bibfnamefont{U.}~\bibnamefont{Baek}},
  \bibinfo{author}{\bibfnamefont{B.}~\bibnamefont{O'Gorman}}, \bibnamefont{and}
  \bibinfo{author}{\bibfnamefont{K.~B.} \bibnamefont{Whaley}},
  \bibinfo{journal}{arXiv:1909.09114 [physics, physics:quant-ph]}
  (\bibinfo{year}{2019}{\natexlab{b}}), \eprint{1909.09114}.

\bibitem[{\citenamefont{Lee et~al.}(2018)\citenamefont{Lee, Huggins,
  {Head-Gordon}, and Whaley}}]{leeGeneralizedUnitaryCoupled2018b}
\bibinfo{author}{\bibfnamefont{J.}~\bibnamefont{Lee}},
  \bibinfo{author}{\bibfnamefont{W.~J.} \bibnamefont{Huggins}},
  \bibinfo{author}{\bibfnamefont{M.}~\bibnamefont{{Head-Gordon}}},
  \bibnamefont{and} \bibinfo{author}{\bibfnamefont{K.~B.} \bibnamefont{Whaley}}
  (\bibinfo{year}{2018}).

\bibitem[{\citenamefont{Babbush et~al.}(2018)\citenamefont{Babbush, Wiebe,
  McClean, McClain, Neven, and Chan}}]{babbush2018low}
\bibinfo{author}{\bibfnamefont{R.}~\bibnamefont{Babbush}},
  \bibinfo{author}{\bibfnamefont{N.}~\bibnamefont{Wiebe}},
  \bibinfo{author}{\bibfnamefont{J.}~\bibnamefont{McClean}},
  \bibinfo{author}{\bibfnamefont{J.}~\bibnamefont{McClain}},
  \bibinfo{author}{\bibfnamefont{H.}~\bibnamefont{Neven}}, \bibnamefont{and}
  \bibinfo{author}{\bibfnamefont{G.~K.-L.} \bibnamefont{Chan}},
  \bibinfo{journal}{Physical Review X} \textbf{\bibinfo{volume}{8}},
  \bibinfo{pages}{011044} (\bibinfo{year}{2018}).

\bibitem[{\citenamefont{Verteletskyi et~al.}(2020)\citenamefont{Verteletskyi,
  Yen, and Izmaylov}}]{verteletskyi2019measurement}
\bibinfo{author}{\bibfnamefont{V.}~\bibnamefont{Verteletskyi}},
  \bibinfo{author}{\bibfnamefont{T.-C.} \bibnamefont{Yen}}, \bibnamefont{and}
  \bibinfo{author}{\bibfnamefont{A.~F.} \bibnamefont{Izmaylov}},
  \bibinfo{journal}{The Journal of Chemical Physics}
  \textbf{\bibinfo{volume}{152}}, \bibinfo{pages}{124114}
  (\bibinfo{year}{2020}), \eprint{https://doi.org/10.1063/1.5141458},
  \urlprefix\url{https://doi.org/10.1063/1.5141458}.

\bibitem[{\citenamefont{Huggins
  et~al.}(2019{\natexlab{c}})\citenamefont{Huggins, McClean, Rubin, Jiang,
  Wiebe, Whaley, and Babbush}}]{hugginsEfficientNoiseResilient2019}
\bibinfo{author}{\bibfnamefont{W.~J.} \bibnamefont{Huggins}},
  \bibinfo{author}{\bibfnamefont{J.}~\bibnamefont{McClean}},
  \bibinfo{author}{\bibfnamefont{N.}~\bibnamefont{Rubin}},
  \bibinfo{author}{\bibfnamefont{Z.}~\bibnamefont{Jiang}},
  \bibinfo{author}{\bibfnamefont{N.}~\bibnamefont{Wiebe}},
  \bibinfo{author}{\bibfnamefont{K.~B.} \bibnamefont{Whaley}},
  \bibnamefont{and} \bibinfo{author}{\bibfnamefont{R.}~\bibnamefont{Babbush}},
  \bibinfo{journal}{arXiv:1907.13117 [physics, physics:quant-ph]}
  (\bibinfo{year}{2019}{\natexlab{c}}), \eprint{1907.13117}.

\bibitem[{\citenamefont{Zhao et~al.}(2020)\citenamefont{Zhao, Tranter, Kirby,
  Ung, Miyake, and Love}}]{PhysRevA.101.062322}
\bibinfo{author}{\bibfnamefont{A.}~\bibnamefont{Zhao}},
  \bibinfo{author}{\bibfnamefont{A.}~\bibnamefont{Tranter}},
  \bibinfo{author}{\bibfnamefont{W.~M.} \bibnamefont{Kirby}},
  \bibinfo{author}{\bibfnamefont{S.~F.} \bibnamefont{Ung}},
  \bibinfo{author}{\bibfnamefont{A.}~\bibnamefont{Miyake}}, \bibnamefont{and}
  \bibinfo{author}{\bibfnamefont{P.~J.} \bibnamefont{Love}},
  \bibinfo{journal}{Phys. Rev. A} \textbf{\bibinfo{volume}{101}},
  \bibinfo{pages}{062322} (\bibinfo{year}{2020}),
  \urlprefix\url{https://link.aps.org/doi/10.1103/PhysRevA.101.062322}.

\bibitem[{\citenamefont{Koch et~al.}(2007)\citenamefont{Koch, Yu, Gambetta,
  Houck, Schuster, Majer, Blais, Devoret, Girvin, and
  Schoelkopf}}]{kochChargeinsensitiveQubitDesign2007}
\bibinfo{author}{\bibfnamefont{J.}~\bibnamefont{Koch}},
  \bibinfo{author}{\bibfnamefont{T.~M.} \bibnamefont{Yu}},
  \bibinfo{author}{\bibfnamefont{J.}~\bibnamefont{Gambetta}},
  \bibinfo{author}{\bibfnamefont{A.~A.} \bibnamefont{Houck}},
  \bibinfo{author}{\bibfnamefont{D.~I.} \bibnamefont{Schuster}},
  \bibinfo{author}{\bibfnamefont{J.}~\bibnamefont{Majer}},
  \bibinfo{author}{\bibfnamefont{A.}~\bibnamefont{Blais}},
  \bibinfo{author}{\bibfnamefont{M.~H.} \bibnamefont{Devoret}},
  \bibinfo{author}{\bibfnamefont{S.~M.} \bibnamefont{Girvin}},
  \bibnamefont{and} \bibinfo{author}{\bibfnamefont{R.~J.}
  \bibnamefont{Schoelkopf}}, \bibinfo{journal}{Phys. Rev. A}
  \textbf{\bibinfo{volume}{76}}, \bibinfo{pages}{042319}
  (\bibinfo{year}{2007}).

\bibitem[{\citenamefont{Deffner and Campbell}(2017)}]{QSL_2017}
\bibinfo{author}{\bibfnamefont{S.}~\bibnamefont{Deffner}} \bibnamefont{and}
  \bibinfo{author}{\bibfnamefont{S.}~\bibnamefont{Campbell}},
  \bibinfo{journal}{J. Phys. A: Math. Theor.} \textbf{\bibinfo{volume}{50}},
  \bibinfo{pages}{453001} (\bibinfo{year}{2017}),
  \urlprefix\url{https://doi.org/10.1088/1751-8121/aa86c6}.

\bibitem[{\citenamefont{Ku et~al.}(2017)\citenamefont{Ku, Long, Wu, Bal, Lake,
  Barnes, Economou, and Pappas}}]{hyperbolic_secant_pulse_pappas}
\bibinfo{author}{\bibfnamefont{H.~S.} \bibnamefont{Ku}},
  \bibinfo{author}{\bibfnamefont{J.~L.} \bibnamefont{Long}},
  \bibinfo{author}{\bibfnamefont{X.}~\bibnamefont{Wu}},
  \bibinfo{author}{\bibfnamefont{M.}~\bibnamefont{Bal}},
  \bibinfo{author}{\bibfnamefont{R.~E.} \bibnamefont{Lake}},
  \bibinfo{author}{\bibfnamefont{E.}~\bibnamefont{Barnes}},
  \bibinfo{author}{\bibfnamefont{S.~E.} \bibnamefont{Economou}},
  \bibnamefont{and} \bibinfo{author}{\bibfnamefont{D.~P.}
  \bibnamefont{Pappas}}, \bibinfo{journal}{Phys. Rev. A}
  \textbf{\bibinfo{volume}{96}}, \bibinfo{pages}{042339}
  (\bibinfo{year}{2017}),
  \urlprefix\url{https://link.aps.org/doi/10.1103/PhysRevA.96.042339}.

\bibitem[{\citenamefont{Choquette et~al.}(2020)\citenamefont{Choquette,
  Di~Paolo, Barkoutsos, S{\'e}n{\'e}chal, Tavernelli, and
  Blais}}]{choquetteQuantumoptimalcontrolinspiredAnsatzVariational2020}
\bibinfo{author}{\bibfnamefont{A.}~\bibnamefont{Choquette}},
  \bibinfo{author}{\bibfnamefont{A.}~\bibnamefont{Di~Paolo}},
  \bibinfo{author}{\bibfnamefont{P.~K.} \bibnamefont{Barkoutsos}},
  \bibinfo{author}{\bibfnamefont{D.}~\bibnamefont{S{\'e}n{\'e}chal}},
  \bibinfo{author}{\bibfnamefont{I.}~\bibnamefont{Tavernelli}},
  \bibnamefont{and} \bibinfo{author}{\bibfnamefont{A.}~\bibnamefont{Blais}},
  \bibinfo{journal}{arXiv:2008.01098 [quant-ph]}  (\bibinfo{year}{2020}),
  \eprint{2008.01098}.

\bibitem[{\citenamefont{Yang et~al.}(2017)\citenamefont{Yang, Rahmani, Shabani,
  Neven, and Chamon}}]{yangOptimizingVariationalQuantum2017}
\bibinfo{author}{\bibfnamefont{Z.-C.} \bibnamefont{Yang}},
  \bibinfo{author}{\bibfnamefont{A.}~\bibnamefont{Rahmani}},
  \bibinfo{author}{\bibfnamefont{A.}~\bibnamefont{Shabani}},
  \bibinfo{author}{\bibfnamefont{H.}~\bibnamefont{Neven}}, \bibnamefont{and}
  \bibinfo{author}{\bibfnamefont{C.}~\bibnamefont{Chamon}},
  \bibinfo{journal}{Phys. Rev. X} \textbf{\bibinfo{volume}{7}},
  \bibinfo{pages}{021027} (\bibinfo{year}{2017}).

\bibitem[{\citenamefont{Abraham et~al.}(2019)\citenamefont{Abraham, AduOffei,
  Akhalwaya, Aleksandrowicz, Alexander, Arbel, Asfaw, Azaustre, AzizNgoueya,
  Barkoutsos et~al.}}]{Qiskit}
\bibinfo{author}{\bibfnamefont{H.}~\bibnamefont{Abraham}},
  \bibinfo{author}{\bibnamefont{AduOffei}},
  \bibinfo{author}{\bibfnamefont{I.~Y.} \bibnamefont{Akhalwaya}},
  \bibinfo{author}{\bibfnamefont{G.}~\bibnamefont{Aleksandrowicz}},
  \bibinfo{author}{\bibfnamefont{T.}~\bibnamefont{Alexander}},
  \bibinfo{author}{\bibfnamefont{E.}~\bibnamefont{Arbel}},
  \bibinfo{author}{\bibfnamefont{A.}~\bibnamefont{Asfaw}},
  \bibinfo{author}{\bibfnamefont{C.}~\bibnamefont{Azaustre}},
  \bibinfo{author}{\bibnamefont{AzizNgoueya}},
  \bibinfo{author}{\bibfnamefont{P.}~\bibnamefont{Barkoutsos}},
  \bibnamefont{et~al.}, \emph{\bibinfo{title}{Qiskit: An open-source framework
  for quantum computing}} (\bibinfo{year}{2019}).

\bibitem[{\citenamefont{Johansson et~al.}(2013)\citenamefont{Johansson, Nation,
  and Nori}}]{qutip}
\bibinfo{author}{\bibfnamefont{J.}~\bibnamefont{Johansson}},
  \bibinfo{author}{\bibfnamefont{P.}~\bibnamefont{Nation}}, \bibnamefont{and}
  \bibinfo{author}{\bibfnamefont{F.}~\bibnamefont{Nori}},
  \bibinfo{journal}{Computer Physics Communications}
  \textbf{\bibinfo{volume}{184}}, \bibinfo{pages}{1234 }
  (\bibinfo{year}{2013}), ISSN \bibinfo{issn}{0010-4655},
  \urlprefix\url{http://www.sciencedirect.com/science/article/pii/S0010465512003955}.

\bibitem[{\citenamefont{Sun et~al.}(2018)\citenamefont{Sun, Berkelbach, Blunt,
  Booth, Guo, Li, Liu, McClain, Sayfutyarova, Sharma et~al.}}]{pyscf}
\bibinfo{author}{\bibfnamefont{Q.}~\bibnamefont{Sun}},
  \bibinfo{author}{\bibfnamefont{T.~C.} \bibnamefont{Berkelbach}},
  \bibinfo{author}{\bibfnamefont{N.~S.} \bibnamefont{Blunt}},
  \bibinfo{author}{\bibfnamefont{G.~H.} \bibnamefont{Booth}},
  \bibinfo{author}{\bibfnamefont{S.}~\bibnamefont{Guo}},
  \bibinfo{author}{\bibfnamefont{Z.}~\bibnamefont{Li}},
  \bibinfo{author}{\bibfnamefont{J.}~\bibnamefont{Liu}},
  \bibinfo{author}{\bibfnamefont{J.~D.} \bibnamefont{McClain}},
  \bibinfo{author}{\bibfnamefont{E.~R.} \bibnamefont{Sayfutyarova}},
  \bibinfo{author}{\bibfnamefont{S.}~\bibnamefont{Sharma}},
  \bibnamefont{et~al.}, \bibinfo{journal}{WIREs Computational Molecular
  Science} \textbf{\bibinfo{volume}{8}}, \bibinfo{pages}{e1340}
  (\bibinfo{year}{2018}),
  \eprint{https://onlinelibrary.wiley.com/doi/pdf/10.1002/wcms.1340},
  \urlprefix\url{https://onlinelibrary.wiley.com/doi/abs/10.1002/wcms.1340}.

\bibitem[{\citenamefont{Wang et~al.}(2015)\citenamefont{Wang, Dolde, Biamonte,
  Babbush, Bergholm, Yang, Jakobi, Neumann, {Aspuru-Guzik}, Whitfield
  et~al.}}]{wangQuantumSimulationHelium2015}
\bibinfo{author}{\bibfnamefont{Y.}~\bibnamefont{Wang}},
  \bibinfo{author}{\bibfnamefont{F.}~\bibnamefont{Dolde}},
  \bibinfo{author}{\bibfnamefont{J.}~\bibnamefont{Biamonte}},
  \bibinfo{author}{\bibfnamefont{R.}~\bibnamefont{Babbush}},
  \bibinfo{author}{\bibfnamefont{V.}~\bibnamefont{Bergholm}},
  \bibinfo{author}{\bibfnamefont{S.}~\bibnamefont{Yang}},
  \bibinfo{author}{\bibfnamefont{I.}~\bibnamefont{Jakobi}},
  \bibinfo{author}{\bibfnamefont{P.}~\bibnamefont{Neumann}},
  \bibinfo{author}{\bibfnamefont{A.}~\bibnamefont{{Aspuru-Guzik}}},
  \bibinfo{author}{\bibfnamefont{J.~D.} \bibnamefont{Whitfield}},
  \bibnamefont{et~al.}, \bibinfo{journal}{ACS Nano}
  \textbf{\bibinfo{volume}{9}}, \bibinfo{pages}{7769} (\bibinfo{year}{2015}),
  ISSN \bibinfo{issn}{1936-0851}.

\bibitem[{\citenamefont{Shen et~al.}(2017)\citenamefont{Shen, Zhang, Zhang,
  Zhang, Yung, and Kim}}]{Shen2017}
\bibinfo{author}{\bibfnamefont{Y.}~\bibnamefont{Shen}},
  \bibinfo{author}{\bibfnamefont{X.}~\bibnamefont{Zhang}},
  \bibinfo{author}{\bibfnamefont{S.}~\bibnamefont{Zhang}},
  \bibinfo{author}{\bibfnamefont{J.-N.} \bibnamefont{Zhang}},
  \bibinfo{author}{\bibfnamefont{M.-H.} \bibnamefont{Yung}}, \bibnamefont{and}
  \bibinfo{author}{\bibfnamefont{K.}~\bibnamefont{Kim}},
  \bibinfo{journal}{Phys. Rev. A} \textbf{\bibinfo{volume}{95}},
  \bibinfo{pages}{020501} (\bibinfo{year}{2017}),
  \urlprefix\url{https://link.aps.org/doi/10.1103/PhysRevA.95.020501}.

\bibitem[{\citenamefont{Colless et~al.}(2018)\citenamefont{Colless, Ramasesh,
  Dahlen, Blok, Kimchi-Schwartz, McClean, Carter, de~Jong, and
  Siddiqi}}]{Colless2018}
\bibinfo{author}{\bibfnamefont{J.~I.} \bibnamefont{Colless}},
  \bibinfo{author}{\bibfnamefont{V.~V.} \bibnamefont{Ramasesh}},
  \bibinfo{author}{\bibfnamefont{D.}~\bibnamefont{Dahlen}},
  \bibinfo{author}{\bibfnamefont{M.~S.} \bibnamefont{Blok}},
  \bibinfo{author}{\bibfnamefont{M.~E.} \bibnamefont{Kimchi-Schwartz}},
  \bibinfo{author}{\bibfnamefont{J.~R.} \bibnamefont{McClean}},
  \bibinfo{author}{\bibfnamefont{J.}~\bibnamefont{Carter}},
  \bibinfo{author}{\bibfnamefont{W.~A.} \bibnamefont{de~Jong}},
  \bibnamefont{and} \bibinfo{author}{\bibfnamefont{I.}~\bibnamefont{Siddiqi}},
  \bibinfo{journal}{Phys. Rev. X} \textbf{\bibinfo{volume}{8}},
  \bibinfo{pages}{011021} (\bibinfo{year}{2018}),
  \urlprefix\url{https://link.aps.org/doi/10.1103/PhysRevX.8.011021}.

\bibitem[{\citenamefont{McCaskey et~al.}(2019)\citenamefont{McCaskey, Parks,
  Jakowski, Moore, Morris, Humble, and Pooser}}]{mccaskey2019quantum}
\bibinfo{author}{\bibfnamefont{A.~J.} \bibnamefont{McCaskey}},
  \bibinfo{author}{\bibfnamefont{Z.~P.} \bibnamefont{Parks}},
  \bibinfo{author}{\bibfnamefont{J.}~\bibnamefont{Jakowski}},
  \bibinfo{author}{\bibfnamefont{S.~V.} \bibnamefont{Moore}},
  \bibinfo{author}{\bibfnamefont{T.~D.} \bibnamefont{Morris}},
  \bibinfo{author}{\bibfnamefont{T.~S.} \bibnamefont{Humble}},
  \bibnamefont{and} \bibinfo{author}{\bibfnamefont{R.~C.}
  \bibnamefont{Pooser}}, \bibinfo{journal}{npj Quantum Information}
  \textbf{\bibinfo{volume}{5}}, \bibinfo{pages}{1} (\bibinfo{year}{2019}), ISSN
  \bibinfo{issn}{2056-6387}.

\bibitem[{\citenamefont{Xia et~al.}(2018)\citenamefont{Xia, Bian, and
  Kais}}]{ishing_kais}
\bibinfo{author}{\bibfnamefont{R.}~\bibnamefont{Xia}},
  \bibinfo{author}{\bibfnamefont{T.}~\bibnamefont{Bian}}, \bibnamefont{and}
  \bibinfo{author}{\bibfnamefont{S.}~\bibnamefont{Kais}}, \bibinfo{journal}{The
  Journal of Physical Chemistry B} \textbf{\bibinfo{volume}{122}},
  \bibinfo{pages}{3384} (\bibinfo{year}{2018}), \bibinfo{note}{pMID: 29099600},
  \eprint{https://doi.org/10.1021/acs.jpcb.7b10371},
  \urlprefix\url{https://doi.org/10.1021/acs.jpcb.7b10371}.

\bibitem[{\citenamefont{Hempel et~al.}(2018)\citenamefont{Hempel, Maier,
  Romero, McClean, Monz, Shen, Jurcevic, Lanyon, Love, Babbush
  et~al.}}]{trapedIon_Roos}
\bibinfo{author}{\bibfnamefont{C.}~\bibnamefont{Hempel}},
  \bibinfo{author}{\bibfnamefont{C.}~\bibnamefont{Maier}},
  \bibinfo{author}{\bibfnamefont{J.}~\bibnamefont{Romero}},
  \bibinfo{author}{\bibfnamefont{J.}~\bibnamefont{McClean}},
  \bibinfo{author}{\bibfnamefont{T.}~\bibnamefont{Monz}},
  \bibinfo{author}{\bibfnamefont{H.}~\bibnamefont{Shen}},
  \bibinfo{author}{\bibfnamefont{P.}~\bibnamefont{Jurcevic}},
  \bibinfo{author}{\bibfnamefont{B.~P.} \bibnamefont{Lanyon}},
  \bibinfo{author}{\bibfnamefont{P.}~\bibnamefont{Love}},
  \bibinfo{author}{\bibfnamefont{R.}~\bibnamefont{Babbush}},
  \bibnamefont{et~al.}, \bibinfo{journal}{Phys. Rev. X}
  \textbf{\bibinfo{volume}{8}}, \bibinfo{pages}{031022} (\bibinfo{year}{2018}),
  \urlprefix\url{https://link.aps.org/doi/10.1103/PhysRevX.8.031022}.

\bibitem[{\citenamefont{Sugisaki et~al.}(2019)\citenamefont{Sugisaki, Nakazawa,
  Toyota, Sato, Shiomi, and Takui}}]{mcqc_takui}
\bibinfo{author}{\bibfnamefont{K.}~\bibnamefont{Sugisaki}},
  \bibinfo{author}{\bibfnamefont{S.}~\bibnamefont{Nakazawa}},
  \bibinfo{author}{\bibfnamefont{K.}~\bibnamefont{Toyota}},
  \bibinfo{author}{\bibfnamefont{K.}~\bibnamefont{Sato}},
  \bibinfo{author}{\bibfnamefont{D.}~\bibnamefont{Shiomi}}, \bibnamefont{and}
  \bibinfo{author}{\bibfnamefont{T.}~\bibnamefont{Takui}},
  \bibinfo{journal}{ACS Central Science} \textbf{\bibinfo{volume}{5}},
  \bibinfo{pages}{167} (\bibinfo{year}{2019}),
  \eprint{https://doi.org/10.1021/acscentsci.8b00788},
  \urlprefix\url{https://doi.org/10.1021/acscentsci.8b00788}.

\bibitem[{\citenamefont{Ritter}(2019)}]{Ritter_2019}
\bibinfo{author}{\bibfnamefont{M.~B.} \bibnamefont{Ritter}},
  \bibinfo{journal}{Journal of Physics: Conference Series}
  \textbf{\bibinfo{volume}{1290}}, \bibinfo{pages}{012003}
  (\bibinfo{year}{2019}),
  \urlprefix\url{https://doi.org/10.1088%2F1742-6596%2F1290%2F1%2F012003}.

\bibitem[{\citenamefont{Sagastizabal et~al.}(2019)\citenamefont{Sagastizabal,
  Bonet-Monroig, Singh, Rol, Bultink, Fu, Price, Ostroukh, Muthusubramanian,
  Bruno et~al.}}]{symm_verf_vqe}
\bibinfo{author}{\bibfnamefont{R.}~\bibnamefont{Sagastizabal}},
  \bibinfo{author}{\bibfnamefont{X.}~\bibnamefont{Bonet-Monroig}},
  \bibinfo{author}{\bibfnamefont{M.}~\bibnamefont{Singh}},
  \bibinfo{author}{\bibfnamefont{M.~A.} \bibnamefont{Rol}},
  \bibinfo{author}{\bibfnamefont{C.~C.} \bibnamefont{Bultink}},
  \bibinfo{author}{\bibfnamefont{X.}~\bibnamefont{Fu}},
  \bibinfo{author}{\bibfnamefont{C.~H.} \bibnamefont{Price}},
  \bibinfo{author}{\bibfnamefont{V.~P.} \bibnamefont{Ostroukh}},
  \bibinfo{author}{\bibfnamefont{N.}~\bibnamefont{Muthusubramanian}},
  \bibinfo{author}{\bibfnamefont{A.}~\bibnamefont{Bruno}},
  \bibnamefont{et~al.}, \bibinfo{journal}{Phys. Rev. A}
  \textbf{\bibinfo{volume}{100}}, \bibinfo{pages}{010302}
  (\bibinfo{year}{2019}),
  \urlprefix\url{https://link.aps.org/doi/10.1103/PhysRevA.100.010302}.

\bibitem[{\citenamefont{Armaos et~al.}(2020)\citenamefont{Armaos, Badounas,
  Deligiannis, and Lianos}}]{compchem_qc_lianos}
\bibinfo{author}{\bibfnamefont{V.}~\bibnamefont{Armaos}},
  \bibinfo{author}{\bibfnamefont{D.~A.} \bibnamefont{Badounas}},
  \bibinfo{author}{\bibfnamefont{P.}~\bibnamefont{Deligiannis}},
  \bibnamefont{and} \bibinfo{author}{\bibfnamefont{K.}~\bibnamefont{Lianos}},
  \bibinfo{journal}{Applied Physics A} \textbf{\bibinfo{volume}{126}},
  \bibinfo{pages}{625} (\bibinfo{year}{2020}).

\bibitem[{\citenamefont{Bravyi et~al.}(2017)\citenamefont{Bravyi, Gambetta,
  Mezzacapo, and Temme}}]{Bravyi_2017}
\bibinfo{author}{\bibfnamefont{S.}~\bibnamefont{Bravyi}},
  \bibinfo{author}{\bibfnamefont{J.~M.} \bibnamefont{Gambetta}},
  \bibinfo{author}{\bibfnamefont{A.}~\bibnamefont{Mezzacapo}},
  \bibnamefont{and} \bibinfo{author}{\bibfnamefont{K.}~\bibnamefont{Temme}},
  \bibinfo{journal}{arXiv:1701.08213 [quant-ph]}  (\bibinfo{year}{2017}).

\bibitem[{\citenamefont{G{\"u}sten et~al.}(2019)\citenamefont{G{\"u}sten,
  Wiesemeyer, Neufeld, Menten, Graf, Jacobs, Klein, Ricken, Risacher, and
  Stutzki}}]{gustenAstrophysicalDetectionHelium2019}
\bibinfo{author}{\bibfnamefont{R.}~\bibnamefont{G{\"u}sten}},
  \bibinfo{author}{\bibfnamefont{H.}~\bibnamefont{Wiesemeyer}},
  \bibinfo{author}{\bibfnamefont{D.}~\bibnamefont{Neufeld}},
  \bibinfo{author}{\bibfnamefont{K.~M.} \bibnamefont{Menten}},
  \bibinfo{author}{\bibfnamefont{U.~U.} \bibnamefont{Graf}},
  \bibinfo{author}{\bibfnamefont{K.}~\bibnamefont{Jacobs}},
  \bibinfo{author}{\bibfnamefont{B.}~\bibnamefont{Klein}},
  \bibinfo{author}{\bibfnamefont{O.}~\bibnamefont{Ricken}},
  \bibinfo{author}{\bibfnamefont{C.}~\bibnamefont{Risacher}}, \bibnamefont{and}
  \bibinfo{author}{\bibfnamefont{J.}~\bibnamefont{Stutzki}},
  \bibinfo{journal}{Nature} \textbf{\bibinfo{volume}{568}},
  \bibinfo{pages}{357} (\bibinfo{year}{2019}), ISSN \bibinfo{issn}{1476-4687}.

\bibitem[{\citenamefont{Khaneja et~al.}(2005)\citenamefont{Khaneja, Reiss,
  Kehlet, Schulte-Herbrüggen, and Glaser}}]{grape}
\bibinfo{author}{\bibfnamefont{N.}~\bibnamefont{Khaneja}},
  \bibinfo{author}{\bibfnamefont{T.}~\bibnamefont{Reiss}},
  \bibinfo{author}{\bibfnamefont{C.}~\bibnamefont{Kehlet}},
  \bibinfo{author}{\bibfnamefont{T.}~\bibnamefont{Schulte-Herbrüggen}},
  \bibnamefont{and} \bibinfo{author}{\bibfnamefont{S.~J.}
  \bibnamefont{Glaser}}, \bibinfo{journal}{Journal of Magnetic Resonance}
  \textbf{\bibinfo{volume}{172}}, \bibinfo{pages}{296 } (\bibinfo{year}{2005}),
  ISSN \bibinfo{issn}{1090-7807},
  \urlprefix\url{http://www.sciencedirect.com/science/article/pii/S1090780704003696}.

\bibitem[{\citenamefont{Gokhale et~al.}(2019)\citenamefont{Gokhale, Ding,
  Propson, Winkler, Leung, Shi, Schuster, Hoffmann, and Chong}}]{Gokhale_2019}
\bibinfo{author}{\bibfnamefont{P.}~\bibnamefont{Gokhale}},
  \bibinfo{author}{\bibfnamefont{Y.}~\bibnamefont{Ding}},
  \bibinfo{author}{\bibfnamefont{T.}~\bibnamefont{Propson}},
  \bibinfo{author}{\bibfnamefont{C.}~\bibnamefont{Winkler}},
  \bibinfo{author}{\bibfnamefont{N.}~\bibnamefont{Leung}},
  \bibinfo{author}{\bibfnamefont{Y.}~\bibnamefont{Shi}},
  \bibinfo{author}{\bibfnamefont{D.~I.} \bibnamefont{Schuster}},
  \bibinfo{author}{\bibfnamefont{H.}~\bibnamefont{Hoffmann}}, \bibnamefont{and}
  \bibinfo{author}{\bibfnamefont{F.~T.} \bibnamefont{Chong}},
  \bibinfo{journal}{Proceedings of the 52nd Annual IEEE/ACM International
  Symposium on Microarchitecture}  (\bibinfo{year}{2019}),
  \urlprefix\url{http://dx.doi.org/10.1145/3352460.3358313}.

\bibitem[{\citenamefont{Boutin et~al.}(2017)\citenamefont{Boutin, Andersen,
  Venkatraman, Ferris, and Blais}}]{openGrape}
\bibinfo{author}{\bibfnamefont{S.}~\bibnamefont{Boutin}},
  \bibinfo{author}{\bibfnamefont{C.~K.} \bibnamefont{Andersen}},
  \bibinfo{author}{\bibfnamefont{J.}~\bibnamefont{Venkatraman}},
  \bibinfo{author}{\bibfnamefont{A.~J.} \bibnamefont{Ferris}},
  \bibnamefont{and} \bibinfo{author}{\bibfnamefont{A.}~\bibnamefont{Blais}},
  \bibinfo{journal}{Phys. Rev. A} \textbf{\bibinfo{volume}{96}},
  \bibinfo{pages}{042315} (\bibinfo{year}{2017}),
  \urlprefix\url{https://link.aps.org/doi/10.1103/PhysRevA.96.042315}.

\bibitem[{\citenamefont{Lu et~al.}(2017)\citenamefont{Lu, Li, Li, Katiyar,
  Park, Jihyun, Guanru, Tao, Hang, Guilu et~al.}}]{bootstrap_quantumcontrol}
\bibinfo{author}{\bibfnamefont{D.}~\bibnamefont{Lu}},
  \bibinfo{author}{\bibfnamefont{K.}~\bibnamefont{Li}},
  \bibinfo{author}{\bibfnamefont{J.}~\bibnamefont{Li}},
  \bibinfo{author}{\bibfnamefont{H.}~\bibnamefont{Katiyar}},
  \bibinfo{author}{\bibfnamefont{A.}~\bibnamefont{Park}},
  \bibinfo{author}{\bibfnamefont{F.}~\bibnamefont{Jihyun}},
  \bibinfo{author}{\bibfnamefont{X.}~\bibnamefont{Guanru}},
  \bibinfo{author}{\bibfnamefont{L.}~\bibnamefont{Tao}},
  \bibinfo{author}{\bibfnamefont{L.}~\bibnamefont{Hang}},
  \bibinfo{author}{\bibfnamefont{B.}~\bibnamefont{Guilu}},
  \bibnamefont{et~al.}, \bibinfo{journal}{npj Quantum Information}
  \textbf{\bibinfo{volume}{3}}, \bibinfo{pages}{45} (\bibinfo{year}{2017}).

\bibitem[{\citenamefont{Gradl et~al.}(2006)\citenamefont{Gradl, Sp{\"o}rl,
  Huckle, Glaser, and Schulte-Herbr{\"u}ggen}}]{parallelOptimalControl}
\bibinfo{author}{\bibfnamefont{T.}~\bibnamefont{Gradl}},
  \bibinfo{author}{\bibfnamefont{A.}~\bibnamefont{Sp{\"o}rl}},
  \bibinfo{author}{\bibfnamefont{T.}~\bibnamefont{Huckle}},
  \bibinfo{author}{\bibfnamefont{S.~J.} \bibnamefont{Glaser}},
  \bibnamefont{and}
  \bibinfo{author}{\bibfnamefont{T.}~\bibnamefont{Schulte-Herbr{\"u}ggen}}, in
  \emph{\bibinfo{booktitle}{Euro-Par 2006 Parallel Processing}}, edited by
  \bibinfo{editor}{\bibfnamefont{W.~E.} \bibnamefont{Nagel}},
  \bibinfo{editor}{\bibfnamefont{W.~V.} \bibnamefont{Walter}},
  \bibnamefont{and} \bibinfo{editor}{\bibfnamefont{W.}~\bibnamefont{Lehner}}
  (\bibinfo{publisher}{Springer Berlin Heidelberg}, \bibinfo{address}{Berlin,
  Heidelberg}, \bibinfo{year}{2006}), pp. \bibinfo{pages}{751--762}, ISBN
  \bibinfo{isbn}{978-3-540-37784-9}.

\bibitem[{\citenamefont{{Cheng} et~al.}(2020)\citenamefont{{Cheng}, {Deng}, and
  {Qia}}}]{AccQOC}
\bibinfo{author}{\bibfnamefont{J.}~\bibnamefont{{Cheng}}},
  \bibinfo{author}{\bibfnamefont{H.}~\bibnamefont{{Deng}}}, \bibnamefont{and}
  \bibinfo{author}{\bibfnamefont{X.}~\bibnamefont{{Qia}}}, in
  \emph{\bibinfo{booktitle}{2020 ACM/IEEE 47th Annual International Symposium
  on Computer Architecture (ISCA)}} (\bibinfo{year}{2020}), pp.
  \bibinfo{pages}{543--555}.

\bibitem[{\citenamefont{Williams}(2010)}]{Williams_2010}
\bibinfo{author}{\bibfnamefont{C.~P.} \bibnamefont{Williams}},
  \emph{\bibinfo{title}{Explorations in Quantum Computing}}
  (\bibinfo{publisher}{Springer}, \bibinfo{year}{2010}).

\bibitem[{\citenamefont{Economou and Barnes}(2015)}]{PhysRevB.91.161405}
\bibinfo{author}{\bibfnamefont{S.~E.} \bibnamefont{Economou}} \bibnamefont{and}
  \bibinfo{author}{\bibfnamefont{E.}~\bibnamefont{Barnes}},
  \bibinfo{journal}{Phys. Rev. B} \textbf{\bibinfo{volume}{91}},
  \bibinfo{pages}{161405} (\bibinfo{year}{2015}),
  \urlprefix\url{https://link.aps.org/doi/10.1103/PhysRevB.91.161405}.

\bibitem[{\citenamefont{Barron et~al.}(2020{\natexlab{b}})\citenamefont{Barron,
  Calderon-Vargas, Long, Pappas, and Economou}}]{PhysRevB.101.054508}
\bibinfo{author}{\bibfnamefont{G.~S.} \bibnamefont{Barron}},
  \bibinfo{author}{\bibfnamefont{F.~A.} \bibnamefont{Calderon-Vargas}},
  \bibinfo{author}{\bibfnamefont{J.}~\bibnamefont{Long}},
  \bibinfo{author}{\bibfnamefont{D.~P.} \bibnamefont{Pappas}},
  \bibnamefont{and} \bibinfo{author}{\bibfnamefont{S.~E.}
  \bibnamefont{Economou}}, \bibinfo{journal}{Phys. Rev. B}
  \textbf{\bibinfo{volume}{101}}, \bibinfo{pages}{054508}
  (\bibinfo{year}{2020}{\natexlab{b}}),
  \urlprefix\url{https://link.aps.org/doi/10.1103/PhysRevB.101.054508}.

\bibitem[{\citenamefont{Ball et~al.}(2020)\citenamefont{Ball, Biercuk,
  Carvalho, Chen, Hush, De~Castro, Li, Liebermann, Slatyer, Edmunds
  et~al.}}]{ballSoftwareToolsQuantum2020}
\bibinfo{author}{\bibfnamefont{H.}~\bibnamefont{Ball}},
  \bibinfo{author}{\bibfnamefont{M.~J.} \bibnamefont{Biercuk}},
  \bibinfo{author}{\bibfnamefont{A.}~\bibnamefont{Carvalho}},
  \bibinfo{author}{\bibfnamefont{J.}~\bibnamefont{Chen}},
  \bibinfo{author}{\bibfnamefont{M.}~\bibnamefont{Hush}},
  \bibinfo{author}{\bibfnamefont{L.~A.} \bibnamefont{De~Castro}},
  \bibinfo{author}{\bibfnamefont{L.}~\bibnamefont{Li}},
  \bibinfo{author}{\bibfnamefont{P.~J.} \bibnamefont{Liebermann}},
  \bibinfo{author}{\bibfnamefont{H.~J.} \bibnamefont{Slatyer}},
  \bibinfo{author}{\bibfnamefont{C.}~\bibnamefont{Edmunds}},
  \bibnamefont{et~al.}, \bibinfo{journal}{arXiv:2001.04060 [quant-ph]}
  (\bibinfo{year}{2020}), \eprint{2001.04060}.

\end{thebibliography}
\end{document}